\documentclass[epj,nopacs]{svjour}
\usepackage{graphics}
\usepackage[italic]{hepnames}
\usepackage[italic]{hepparticles}
\usepackage{units}
\usepackage{upgreek,amssymb}
\usepackage{array,multirow,graphicx}
\usepackage{float}
\usepackage{url}

\RequirePackage[colorinlistoftodos,prependcaption,textsize=scriptsize,backgroundcolor=orange!20]{todonotes}

\newcommand{\dd}{\ensuremath{\mathrm d}}
\newcommand{\er}{$\pm$}
\newcommand{\ph}{\phantom}
\newcommand{\hs}{\hspace{-4mm}}
\newcommand{\hf}{\hfill}
\newcommand{\hp}{\hf\ph}
\newcommand{\ty}{\tiny}
\newcommand{\LlJ}{\hs$L$$<$$J$\hs}
\newcommand{\LgJ}{\hs$L$$>$$J$\hs}

\begin{document}
\title{Polarization observables in double neutral pion photoproduction}
\titlerunning{Polarization observables in double neutral pion photoproduction}
\authorrunning{The CBELSA/TAPS Collaboration}
\author{The CBELSA/TAPS Collaboration\medskip \\  
 T.~Seifen\inst{1},
 J.~Hartmann\inst{1},
 F.~Afzal\inst{1},
 A.V.~Anisovich\inst{1},
 R.~Beck\inst{1},
 M.~Becker\inst{1},
 A.~Berlin\inst{5},
 M.~Bichow\inst{5},
 \mbox{K.-Th.~Brinkmann\inst{1,3},}
 V.~Crede\inst{6},
 M.~Dieterle\inst{4},
 H.~Dutz\inst{2},
 H.~Eberhardt\inst{2},
 D.~Elsner\inst{2},
 K.~Fornet-Ponse\inst{2},
 St.~Friedrich\inst{3},
 F.~Frommberger\inst{2},
 Ch.~Funke\inst{1},
 M.~Gottschall\inst{1},
 M.~Gr\"uner\inst{1},
 St.~G\"ortz\inst{2},
 E.~Gutz\inst{1,3},
 Ch.~Hammann\inst{1},
 J.~Hannappel\inst{2},
 J.~Herick\inst{5},
 W.~Hillert\inst{2,}\thanks{\emph{Present address:} Universit\"{a}t Hamburg, Germany},
 Ph.~Hoffmeister\inst{1},
 Ch.~Honisch\inst{1},
 O.~Jahn\inst{2},
 T.~Jude\inst{2},
 A.~K\"aser\inst{4},
 D.~Kaiser\inst{1},
 H.~Kalinowsky\inst{1},
 F.~Kalischewski\inst{1},
 P.~Klassen\inst{1},
 I.~Keshelashvili\inst{4},
 F.~Klein\inst{2},
 E.~Klempt\inst{1},
 K.~Koop\inst{1},
 B.~Krusche\inst{4,}\thanks{deceased},
 M.~Lang\inst{1},
 Ph.~Mahlberg\inst{1},
 K.~Makonyi\inst{3},
 F.~Messi\inst{2,}\thanks{\emph{Present address:} Lunds universitet, Sweden},
 V.~Metag\inst{3},
 W.~Meyer\inst{5},
 J.~M\"uller\inst{1},
 J.~M\"ullers\inst{1},
 M.~Nanova\inst{3},
 K.~Nikonov\inst{1},
 V.A.~Nikonov\inst{1,\rm b},
 R.~Novotny\inst{3},
 S.~Reeve\inst{2},
 B.~Roth\inst{5},
 G.~Reicherz\inst{5},
 T.~Rostomyan\inst{4},
 St.~Runkel\inst{2},
 A.V.~Sarantsev\inst{1},
 Ch.~Schmidt\inst{1},
 H.~Schmieden\inst{2},
 R.~Schmitz\inst{1},
 J.~Schultes\inst{1},
 V.~Sokhoyan\inst{1,}\thanks{\emph{Present address:} Universit\"{a}t Mainz, Germany},
 N.~Stausberg\inst{1},
 A.~Thiel\inst{1},
 U.~Thoma\inst{1},
 M.~Urban\inst{1},
 G.~Urff\inst{1},
 H.~van~Pee\inst{1},
 D.~Walther\inst{1},
 Ch.~Wendel\inst{1},
 U.~Wiedner\inst{5},
 A.~Wilson\inst{1,6},
 L.~Witthauer\inst{4}~and
 Y.~Wunderlich\inst{1}
 .}

 \mail{seifen{@}hiskp.uni-bonn.de (T.~Seifen), thoma{@}hiskp.uni-bonn.de (U.~Thoma)}

\institute{
Helmholtz--Institut f\"ur Strahlen-- und Kernphysik, Universit\"at Bonn, 53115 Bonn, Germany \and
Physikalisches Institut, Universit\"at Bonn, 53115 Bonn, Germany \and
Physikalisches Institut, Universit\"at Gie{\ss}en, 35392 Gie{\ss}en, Germany \and
Physikalisches Institut, Universit\"at Basel, 4056 Basel, Switzerland \and
Institut f\"ur Experimentalphysik I, Ruhr--Universit\"at Bochum, 44780  Bochum, Germany \and
Department of Physics, Florida State University, Tallahassee, FL 32306, USA
}

\date{Received: \today / Revised version:}

\abstract{Measurements of target asymmetries and double-polarization observables for the reaction
\mbox{$\Pgg\Pp\to\Pp\Pgpz\Pgpz$} are reported. The data were taken with the CBELSA/TAPS experiment at the
ELSA facility (Bonn University) using the Bonn frozen-spin butanol (C$_4$H$_9$OH) target, which provided
transversely polarized protons. Linearly polarized photons were produced via bremsstrahlung off a
diamond crystal. The data cover the photon energy range from $E_{\Pgg}=\unit[650]{MeV}$ to $E_{\Pgg}=\unit[2600]{MeV}$
and nearly the complete angular range. The results have been included in the BnGa partial wave analysis.
Experimental results and the fit agree very well.
Observed systematic differences in the branching ratios for decays of $\PN^*$ and $\PgD^*$ resonances are attributed to the internal structure of these excited nucleon states.
Resonances which can be assigned to SU(6)$\times$O(3) two-oscillator configurations show larger branching ratios to intermediate states
with non-zero intrinsic orbital angular momenta
than resonances assigned to one-oscillator configurations.
  }

\maketitle
\section{Introduction}
\label{sec:intro}
Studying the excitation spectrum of the nucleon is an important tool to gain a better understanding of the non-perturbative regime of QCD.
Final states originating from the successive decay of intermediate resonances play an important role in understanding the baryon
spectrum~\cite{Capstick:2000qj}. At higher excitation energies ($W\gtrsim\unit[1.9]{GeV}$), where the so-called {\it missing resonances}
are expected, the neutral multi-meson cross sections for photo-induced reactions exceed those
for the production of single neutral mesons. This emphasizes the importance of multi-meson final states. \\
 The {\it missing resonances} are states which are predicted by quark
 models~\cite{Capstick:1986bm,Loring:2001kx} but have not (yet) been found
 experimentally. Whether these ``missing'' states do indeed not exist or
 whether there is a reason for them not to occur in the reactions investigated
 so far, is an important question to be answered. 
 Of course, the quark model with its assumption that just three constituent
 quarks drive the spectrum of baryon resonances may also
 be using the wrong degrees of freedom. 
 One possible explanation, suggested early on, is that instead of three
 constituent quarks participating in the internal baryon dynamics,
 two quarks cluster in a diquark, and the quark-diquark system forms an
 underlying structure (for a review on those models see \textit{e.g.}
 Ref.~\cite{anselmino}). This would reduce the number of degrees of
 freedom and the number of resonances. Resonances like the 4-star
 N(1900)$3/2^+$~\cite{RPP,Anisovich:2017bsk} found and confirmed in
 recent years are at variance with such a quark-diquark picture.  
 Interestingly enough, calculations of the baryon spectrum within lattice
 QCD~\cite{Edwards:2011jj,Khan:2020ahz} indicate a level counting consistent
 with the non-relativistic quark model~\cite{Edwards:2011jj}. Of course,
 also these results are based on approximations of QCD. 
A complementary ansatz for the calculation of non-perturbative
QCD-phenomena is the Dyson-Schwinger/Bethe-Salpeter approach, which
also provides interesting results on the spectrum of non-strange and
strange baryon resonances (see~\cite{Eichmann:2016yit,Eichmann:2018adq,Qin:2019hgk,Barabanov:2020jvn} and references therein).
The existing tools allow to solve the covariant three-body
Fadeev-equation or to use certain approximations and treat baryons
as quark-diquark systems, both versions being based on the same underlying
quark-gluon interaction. 
Most calculations in the baryon sector and especially of the baryon spectrum have been performed in rainbow-ladder approximation. 
The spectra of baryon resonances have been calculated for $J=1/2^\pm$ and
$J=3/2^\pm$, reaching for the $N^*$- and $\Delta^*$-resonances to masses
of about 1950\,MeV.   

Alternative approaches to the physics of strong interactions generate hadron resonances from the interaction of their decay products, for details see~\cite{Lutz:2005ip,Guo:2017jvc,Mai:2022eur} and references therein. The hadrogenesis conjecture~\cite{Lutz:2005ip} suggests that it should be
possible to generate the full spectrum of excited hadrons from hadrons of lower masses. Presently, it is still unknown, whether the still {\it missing resonances} of the quark model would also be predicted as part of the spectrum of dynamically generated resonances. 
They may be absent, thus providing a different excitation spectrum to be compared with experiment. 
Yet in any case, an experimental search for new resonances seems rewarding given our present knowledge on the baryon spectrum, also discussed in \cite{Crede:2013kia,Thiel:2022xtb,EK2025}.

A very different interpretation of the high-mass spectrum of baryon resonances was given by
Glozman~\cite{Glozman:1999tk}, who started from the observation that many baryon resonances form
parity doublets, \textit{i.e.} pairs of resonances with similar masses, same total angular momentum $J$ but
opposite parities. At low masses, this symmetry is heavily broken: the nucleon with $J^P=1/2^+$ is
about \unit[600]{MeV} lighter than its parity partner $\PN(1535)1/2^-$. Obviously, chiral symmetry is
broken. Glozman argues that chiral symmetry breaking might not play a role in the high-mass spectrum of hadron
resonances. Then, all resonances should have parity partners. Possibly, an extended symmetry even leads
to mass-degenerate quartets of resonances of $\PN^*$'s and $\PgD^*$'s with opposite parities like
$\PN(1900)3/2^+$, $\PN(1895)3/2^-$, $\PgD(1920)3/2^+$, and $\PgD(1930)3/2^-$. The consequences of
this conjecture were discussed intensely, see reviews
\cite{Jaffe:2004ph,Glozman:2007ek,Afonin:2007mj,Shifman:2007xn}. In \cite{Klempt:2002tt}
it is pointed out that spin-parity partners naturally emerge when the (squared) masses increase with
$M^2\propto L+N$ ($L$ is the total orbital angular momentum, $N$ the radial excitation quantum
number) -- as predicted within AdS/QCD \cite{Forkel:2008un,Brodsky:2014yha} and an empirical mass
formula~\cite{Klempt:2002vp} -- and not with $M^2\propto L+2N$ which follows from the harmonic
oscillator potential. In this case, resonances with $J=L+3/2$ on the leading Regge trajectory (like
$\PgD(1232)3/2^+$, $\PgD(1950)7/2^+$, $\PgD(2420)11/2^+$, $\dots$), should have no parity
partners. Indeed, a search for $\PgD^*$ states with $J^P = 7/2^\pm$ in the 1900 to \unit[2200]{MeV} region
found a positive-parity resonance at \unit[(1917$\pm$4)]{MeV} and a negative-parity resonance at
\unit[(2176$\pm$40)]{MeV}. No negative parity partner of the $\PgD(1950)7/2^+$ was observed \cite{Anisovich:2015gia}. A confirmation of the result and an extension to $\PN^*$'s in the fourth resonance region seem promising. 

When, in a quark-model based on three constituent quarks (e.g.~\cite{Loring:2001kx}), the wave function of an excited baryon is expanded into wave functions of a harmonic
oscillator (with two oscillators $\lambda$ and $\rho$), three classes of wave functions result.
One class has wave functions in which the excitation energy oscillates between
the two oscillators. For an excitation with two units of orbital angular momentum, either
$l_\rho=2, l_\lambda=0$ or $l_\rho=0, l_\lambda=2$. In a second class of wave functions both
oscillators are simultaneously excited, $l_\rho=1, l_\lambda=1$. The majority of resonances belongs
to a third class in which the wave function is given by the coherent sum of both components. 
Detailed calculations show that the harmonic-oscillator wave function with defined orbital angular momentum and total quark spin is
often the leading part of the full wave function with only small admixtures of higher-order
terms~\cite{Loring:2001kx}. 
Therefore, the classes of harmonic-oscillator wave functions correspond
to classes of baryon resonances. Recent studies on double-pion photoproduction
\cite{Sokhoyan:2015fra,Thiel:2015xba} have indicated that the first class of resonances decays preferentially into a
ground-state baryon and a ground-state meson, \textit{e.g.} $\PN\Pgp$ or $\PgD(1232)\Pgp$.
The third class of resonances based on a mixture of single- and two-oscillator excitations also decays into $\PN\Pgp$ or
$\PgD(1232)\Pgp$, but decays were observed in addition in which one particle was an excited hadron: Decays
like those into $\PN(1520)\Pgp$ or $\PN\sigma$ were seen~\cite{Sokhoyan:2015fra}. This was interpreted
as evidence that the component in the wave function in which both oscillators are excited simultaneously, first
de-excites into an excited hadron and a ground-state hadron. In this step, one of the two excited
oscillators de-excites while the other oscillator remains excited. In the final step, the second
oscillator releases its energy as well, so that finally three ground-state hadrons are
produced~\cite{Sokhoyan:2015fra,Thiel:2015xba}.

This interpretation may explain why some of the expected resonances might be missing:
Resonances of the second class, with both oscillators being simultaneously excited, should then decay
only in two steps, never in a single decay process. Hence, they also cannot be produced in a single
step process: They are neither produced in photo- nor in pion-induced reactions. Still, mixing of
states could nevertheless lead to a small production rate but mostly, the calculated mixing angles
are small~\cite{Loring:2001kx}, and the production strength of these resonances is expected to be
small.

The study of sequential decays of high mass resonances into multiple mesons thus offers three
chances, i) to search for {\it missing resonances}, ii) to study parity doublets in the high-mass
region, and iii) to study the internal structure of baryons through the dynamics of sequential
decays. Clearly, high statistics is required for further progress and, even more importantly, the
measurement of new polarization observables.

In this paper, new data on polarization observables for the reaction
\begin{equation}
\Pgg\Pp\to\Pp\Pgpz\Pgpz ,
\label{reaction}
\end{equation}
taken with a transversely polarized target and a linearly polarized photon beam, are presented.
The photoproduction of two neutral pions is particularly well suited to study sequential decays.
It suffers much less from non-resonant contributions than the production of $\Pgpp\Pgpm$
pairs. In reaction~(\ref{reaction}), there is no diffractive $\Pgr$ production, no $\PgD^{++}$-Kroll-Ruderman
term (direct $\PgD^{++}\Pgpm$ production), and contributions from $t$-channel processes or Born terms are
suppressed compared to \textit{e.g.} $\Pgpp\Pgpm$ photoproduction. The $\Pp\Pgpz\Pgpz$-channel is hence
very well suited for the investigation of baryon resonances.

The paper is organized as follows: In Section~\ref{sec:setup}, a short description of the
CBELSA/TAPS experiment is given. In Section~\ref{sec:selection}, the selection criteria are
discussed to obtain a nearly background-free sample of events for reaction~(\ref{reaction}).
Section~\ref{sec:analysis} is devoted to the determination of the polarization observables. These new
data are incorporated into the database for the latest BnGa multichannel partial wave analysis. In
Section~\ref{sec:PWA}, the results of the partial wave analysis are presented. A summary is given in
Section~\ref{sec:summary}.

\section{Experimental setup}
\label{sec:setup}
\begin{figure}
\resizebox{.5\textwidth}{!}{
  \includegraphics{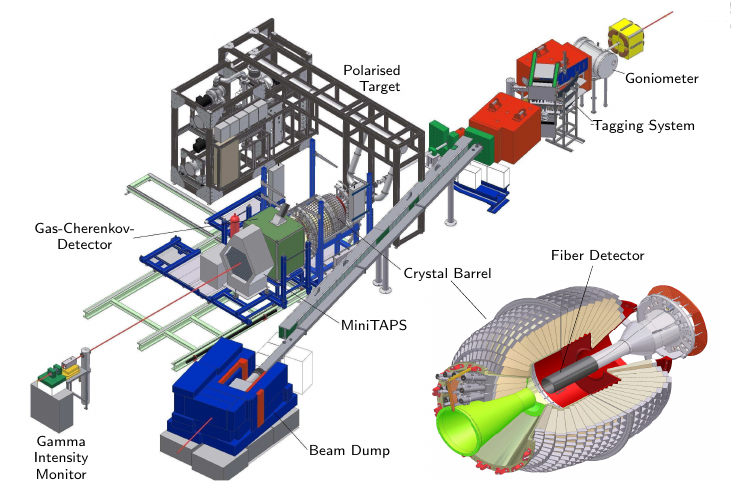}
} \caption{Setup of the CBELSA/TAPS experiment.}
\label{fig:cbexp}
\end{figure}

The CBELSA/TAPS experiment (setup shown in Fig.~\ref{fig:cbexp}) is located at the Electron
Stretcher Accelerator ELSA \cite{elsa} at the University of Bonn. Electrons with an energy of \unit[3.2]{GeV}
impinged on a diamond crystal (thickness $\unit[500]{\upmu m}$) and produced linearly polarized
photons with a maximal polarization of \unit[66]{\%} at \unit[850]{MeV} via coherent brems\-strahlung
\cite{elsner}. The deflection of the electrons in a magnetic dipole field was measured with a tagging system
consisting out of 480 scintillating fibers and 96 scintillating bars, covering the energy range
between \unit[2]{\%} and \unit[88]{\%} of the incident electron energies.

The tagged photons hit the Bonn frozen-spin target \cite{bradtke} placed at the center of the
Crystal Barrel calorimeter. The target material was butanol (C$_4$H$_9$OH) and the protons in the
hydrogen nuclei were transversely polarized. During the data taking the \unit[2]{cm} long target cell
was cooled to \unit[50--60]{mK}. A holding coil produced a magnetic field of about \unit[0.6]{T}.
Relaxation times of several hundred hours were reached with this so-called frozen-spin technique.
Every two to three days a re-polarization of the target was necessary. This lead to a mean polarization of about \unit[74]{\%}.

For background studies the target cell containing the butanol was replaced with a carbon foam
target. Using the same cryostat (then operating at about \unit[1]{K}) and the carbon foam target at
about the same area density as the bound nuclei in the butanol, ensured that the systematic effects
in the data taking and in the background study were negligible.

The Crystal Barrel calorimeter (CB) \cite{cb}, consisting of 1320 CsI(Tl) crystals, is ideally
suited for the detection of photons from the decay of neutral mesons and covered the polar angular range from $11.2^\circ$ -- $156^\circ$. For the three most forward rings of in total 90 crystals ($11.2^\circ$ to $27.5^\circ$), the earlier photodiode read-out has been replaced with
photomultipliers. The TAPS calorimeter \cite{novotny,gabler}, which consists of 216 $\text{BaF}_2$ crystals, was positioned
\unit[210]{cm} away from the target center and covered the lowest polar angle range from $12^\circ$
down to about $1^\circ$. Combined, the two calorimeters covered the polar angle range of $1^\circ$ to
$156^\circ$ and the full azimuthal angle $\varphi$, thus covering about
\unit[95]{\%} of the full solid angle in the laboratory frame.

Charged-particle identification was possible by a scintillating fiber detector (513 fibers in three
layers) \cite{suft} directly surrounding the target, 180 plastic scintillation counters in two layers in front of
the first 90 CB, and plastic scintillators in front of the TAPS crystals.

A CO$_2$-Cherenkov detector was placed between the CB and TAPS calorimeters to identify and suppress
electromagnetic background at the trigger level.

The flux of incoming photons was measured by taking into account the coincidence between the
Gamma-Intensity-Monitor (GIM), a $4\times4$ lead glass matrix at the end of the beam line, and the
tagging system. Since the GIM efficiency decreased at high rates (${\gg\unit[1]{MHz}}$), the FluMo detector, placed right in front of the GIM,
measured $\Pep\Pem$-pairs from photon conversion in a lead foil at low rate, thus allowing to monitor the rate-dependent GIM efficiency.

Trigger conditions demanded at least two energy deposits in the calorimeters and no veto signal from the cherenkov detector. If there was no energy deposit in the TAPS calorimeter or the first 90 crystals of the CB, a charged hit in the inner detector was required since the signals of the central part of the Crystal Barrel calorimeter could not be used in the first-level trigger. In this case, two energy deposits in the CB were required, identified by a fast cluster encoder in the second-level trigger. In case of only one energy deposit below $27.5^\circ$, at least one additional hit was required in the second-level trigger.

For further details on the experimental setup see Ref.~\cite{gottschall}.

\section{Data selection}
\label{sec:selection}
Both pions in the reaction $\Pgg\Pp\to\Pp\Pgpz\Pgpz$ predominantly decay into two photons, leading to four photons in the final state. Therefore,
events with exactly one charged and four neutral detector hits were selected. The charged hit
was then treated as a proton. Hits in the charge sensitive fiber detector or the CB-scintillators in forward direction alone without a
corresponding calorimeter hit were considered proton candidates as well. This was necessary to also
reconstruct protons with low kinetic energy, which did not deposit sufficient energy in the
calorimeters or hit an insensitive region between the calorimeter modules.

\begin{figure*}
\resizebox{\textwidth}{!}{%
  \includegraphics{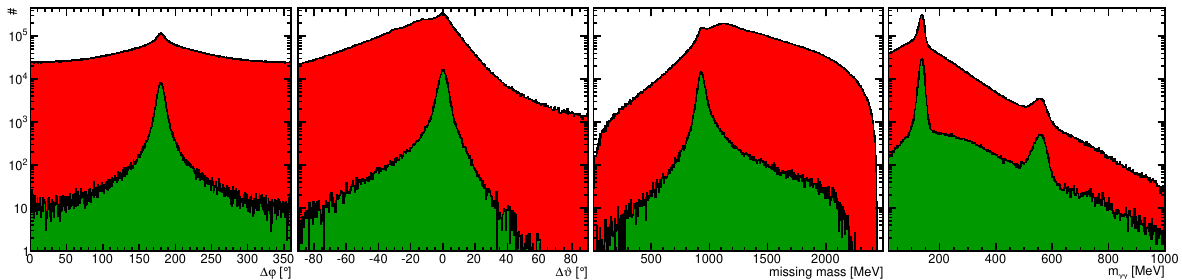}
} \caption{Distributions of the variables used for the kinematic cuts are shown for $\unit[650]{MeV} < E_{\Pgg} < \unit[2600]{MeV}$ on a logarithmic scale. For each of the four cuts (coplanarity, polar angle of the proton, missing mass, and $2\Pgg$-invariant mass) the spectrum is shown in red after just the time cut (including the side-band subtraction) and in green after all cuts except the one on the variable shown. The invariant $2\Pgg$-mass contains all $2\Pgg$ combinations and therefore also combinatorial background.}
\label{fig:cuts}
\end{figure*}

A time coincidence was required between the measured final state particles and the scattered electron in the tagger. 
Random time background in the tagger was subtracted by means of a side-band subtraction. The signal region was between \unit[-16]{ns} and \unit[4]{ns}, allowing for slow protons to reach the TAPS detector. The side-band consisted of two \unit[200]{ns} wide regions, one below and one above the signal region.

\begin{figure}[pb]
\resizebox{0.45\textwidth}{!}{
  \includegraphics{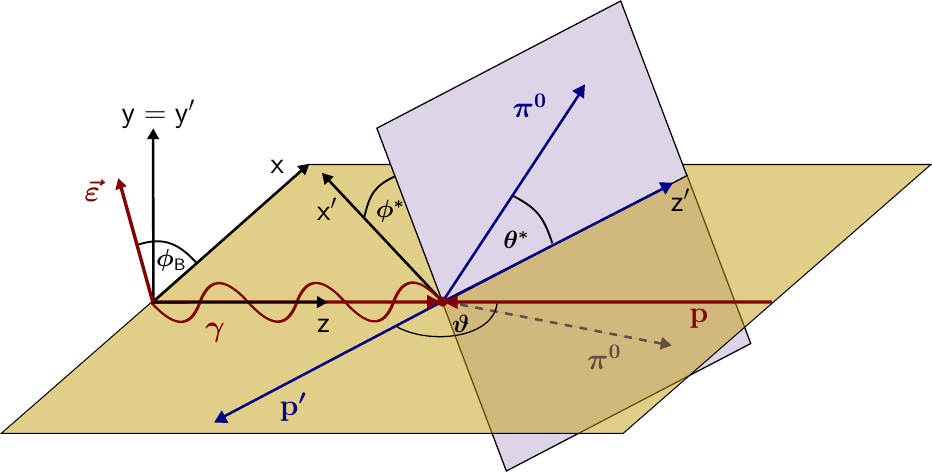}
} \caption{Definition of kinematic variables in the CMS. Details see text.}
\label{fig:3body}
\end{figure}

The kinematics of the reaction $\Pgg\Pp\to\Pp\Pgpz\Pgpz$ can be described by five independent kinematic
variables. One choice of these variables makes use of two planes inherent to the process: the
reaction plane spanned by the incoming photon and one of the outgoing particles $p_1$, and the
decay plane spanned by the other two outgoing particles $p_2$ and $p_3$, as illustrated in
Fig.~\ref{fig:3body}. The kinematic variables then are the energy of the incoming photon
$E_{\Pgg}$, the scattering angle in the CMS $\cos\vartheta_{1}$\footnote{In case of the proton as the recoiling particle, \textit{i.e.} $p_1=\Pp$, $\cos\vartheta_{\Pgpz\Pgpz}=-\cos\vartheta_{1}$ is used instead.}, the invariant mass $m_{23}$ of the particles
$p_2$ and $p_3$, the angle $\phi^*_{23}$ between the reaction plane and the decay plane, and the angle
$\theta^*_{23}$ between the $z'$-axis and the particle $p_2$.

Since the two pions are indistinguishable, two choices for the particles $p_1(p_2p_3)$ remain, namely: $\Pp(\Pgpz\Pgpz)$ and $\Pgpz(\Pp\Pgpz)$.

\subsection{Kinematic cuts}
The kinematics of the initial state of the reaction were known (photon of known energy along the
$z$-axis, target proton at rest). Thus, the four-momentum of one final-state particle could be calculated using energy and momentum conservation.

In order to suppress background events, several kinematic constraints had to be fulfilled for an
event to be retained. In the center-of-mass system (CMS) the final-state proton and the four-photon system
depart back-to-back. Therefore, the difference of the azimuthal angle $\varphi$ of the
proton to the $4\Pgg$-system should be $180^\circ$ (coplanarity). The polar angle $\vartheta$ is influenced by the
Lorentz boost to the laboratory frame. Hence, the difference of the proton hit to the
calculated proton angles was used which should vanish. A Gaussian
distribution was fitted to the angular distributions to determine the cut limits and events within $2\sigma$ were retained.
Additionally, the mass of the calculated proton has to agree with the nominal proton mass. Here, a
Novosibirsk function~\cite{novosibirsk} was fitted and events within the $\unit[2.27]{\%}$ and $\unit[97.73]{\%}$
quantiles were selected, thus retaining $\unit[95.45]{\%}$ of the events just like with the
$\pm2\sigma$ cut in case of the angular differences. The four photons in the final state were grouped pair-wise
to form the two pions. A Novosibirsk function was fitted to both pion mass spectra and events were selected within the same quantiles mentioned above. Fig.~\ref{fig:cuts} shows the
four cut variables on a logarithmic scale before the kinematic cuts were applied and after all but the cut on the shown variable.

The width of the distributions varied with the kinematic variables, \textit{e.g.} the width of the missing mass depended on the energy of the incoming photon. Hence, all the above-mentioned
fits were done separately for different beam energies and polar angles of the $4\Pgg$-system.
Approximate limits for the cuts are given in Table~\ref{tab:cuts}.

\begin{table}
\caption{Approximate limits of the kinematic cuts.}
\label{tab:cuts}
\centering
\begin{tabular}{ccc}
\hline\noalign{\smallskip}
variable & lower limit & upper limit \\
\noalign{\smallskip}\hline\noalign{\smallskip}
$\Delta\varphi$ & $168^\circ\dots175^\circ$ & $185^\circ\dots192^\circ$ \\
$\Delta\vartheta$ & $-11^\circ {\dots} -2^\circ$ & $2^\circ\dots10^\circ$ \\
$m_{\Pp,\text{miss}}$ & $\unit[840]{MeV}\dots\unit[910]{MeV}$ & $\unit[990]{MeV}\dots\unit[1140]{MeV}$ \\
$m_{\Pgg\Pgg}$ & $\unit[105]{MeV}\dots\unit[120]{MeV}$ & $\unit[145]{MeV}\dots\unit[155]{MeV}$ \\
\noalign{\smallskip}\hline
\end{tabular}
\end{table}

\begin{figure}[pb]
\resizebox{0.5\textwidth}{!}{%
  \includegraphics{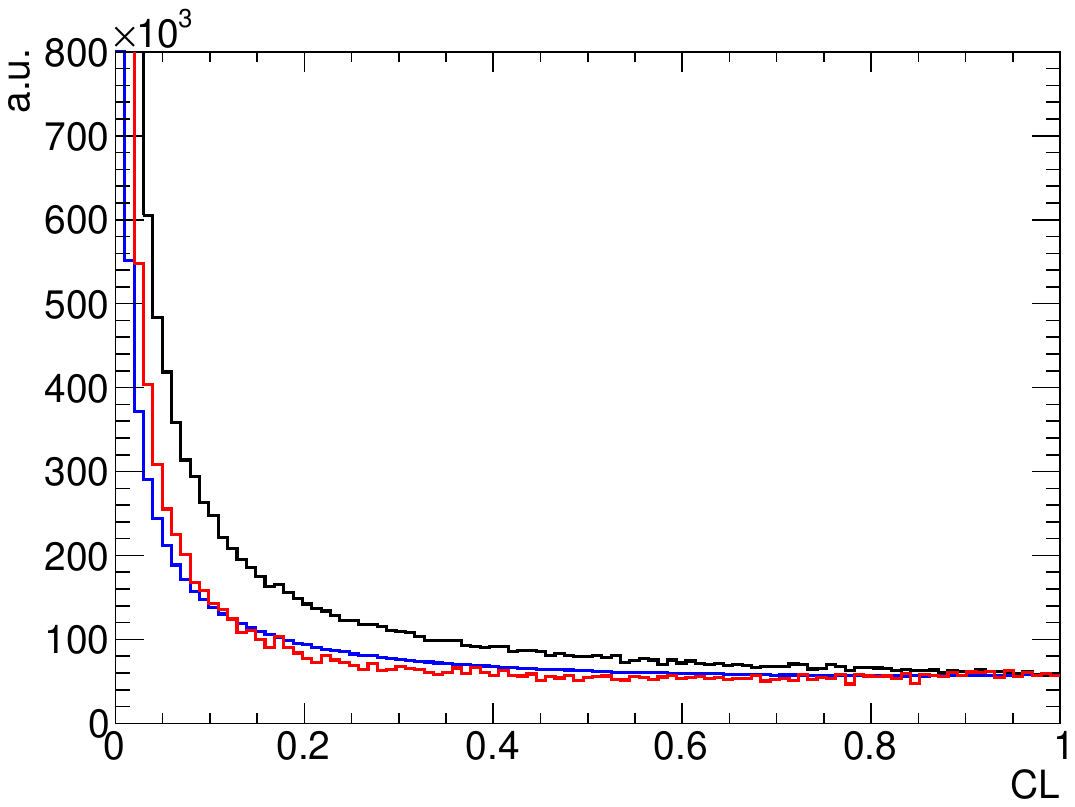}
} \caption{Confidence level distribution of hydrogen data (blue), butanol data (black) and butanol with carbon data subtracted (red) for the
hypothesis $\Pgg\Pp\to\Pp_\text{miss}\Pgpz\Pgpz$. In this plot, only very broad data selection cuts were applied to the data.}
\label{fig:CL}
\end{figure}

\subsection{Kinematic fit}
After applying these kinematic cuts combinatorial background remained, stemming from the three possibilities to combine four photons into two pions. Of
course, mostly only one of those combinations survived the mass cuts but in about \unit[10]{\%} of the cases there were more than one.
A kinematic fit chooses the combination with the highest confidence level (see below), thus eliminating the combinatorial background.
The control spectra of the kinematic fit (pull distribution and confidence level) were determined from a very broad cut event sample ($\gtrsim 5\sigma$).
A detailed description of the kinematic fit can be found in \cite{pee}. Due to the finite detector resolution, energy and momentum conservation were
not fulfilled exactly in the data. The concept of the fit is to vary the measured parameters within their uncertainties to exactly fulfill the constraints of the fit.
Energy and momentum conservation as well as the masses of the final state particles can be used as
constraints. Since the proton was not always stopped in the calorimeters and the energy deposit of protons differs from that of photons, the proton's measured energy did not correspond to its true energy.
Hence, the proton was not fitted but treated as a missing particle and only the three mass
constraints remained for the fitted hypothesis $\Pgg\Pp\to\Pp_\text{miss}\Pgpz\Pgpz$. The
kinematic fit provides a handle on the systematic effects, the so-called pull distribution, which is
the difference of the fitted to the measured variables normalized by the uncertainties of the difference.
The pull distributions  of the uncut events should follow a standard normal distribution (vanishing
mean and unit standard deviation).

An additional criterion for the quality of the fit is the so-called confidence level (CL) which is
the integral over the $\chi^2$ probability density distribution starting at the given $\chi^2$. For independent parameters with normally distributed statistical uncertainties, the CL-distribution should
be flat for events which fulfill the hypothesis (no background). A rise to low values of the CL is due to background
events where the fit has a large $\chi^2$.

Apart from polarizable protons in the hydrogen nuclei, the butanol data also contains events on protons bound in carbon and oxygen nuclei which
were subject to initial Fermi motion. Thus, the hypothesis of an initial proton at rest led to a
steady rise to low values in the CL for those events (\textit{cf.} black line in
Fig.~\ref{fig:CL}). For control purposes, the CL of the kinematic fit was investigated with data
taken on a \unit[5]{cm} long hydrogen target. The resulting CL distribution is shown in Fig.~\ref{fig:CL} as a blue line and shows
the expected flat behavior with the rise at small CL values. Additionally, the CL distribution is shown for butanol events from which the contributing carbon events have been subtracted. Also here the expected flat behavior is observed.
A cut on the CL of $\geq0.1$ was applied to reduce the number of background events in the selected data.

The reaction $\Pgg\Pp\to\Pp\Pgpz\Pgh$ has the same 4\Pgg{} final state and therefore, it is possible for a
photon combination from such a reaction to mimic the two pion reaction. The
hypothesis $\Pgg\Pp\to\Pp_\text{miss}\Pgpz\Pgh$ was also fitted and events which had a higher CL for
that anti-hypothesis than for the $\Pgg\Pp\to\Pp_\text{miss}\Pgpz\Pgpz$ hypothesis were discarded.

After the cuts described above, the selected data sample contained about 254000 events in the energy range of \unit[650--2600]{MeV} shown in section~\ref{sec:results}.

\begin{figure}[pt]
\resizebox{0.5\textwidth}{!}{%
  \includegraphics{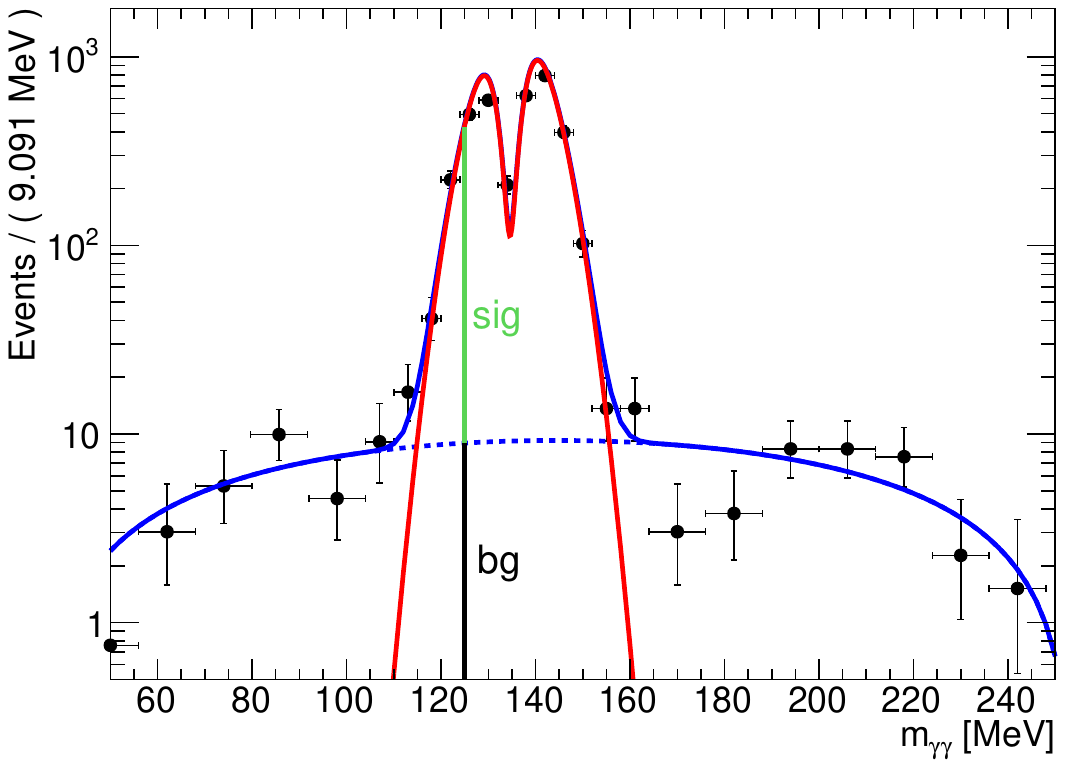}
} \caption{Example of the mass distribution of the nearest neighbors. The signal (red) is described
by the sum of two Gaussian functions (the narrower one with negative amplitude), the background
function (dashed blue) is a second order polynomial. The dip on the pion peak is due to the fact that the fit uses
	the best 2\Pgg{} combination for the fitted pion. The data is binned here for presentational purposes only. The horizontal bars indicate the (variable) bin width.}
\label{fig:bg_fit}
\end{figure}

\begin{figure*}
\resizebox{\textwidth}{!}{%
  \includegraphics{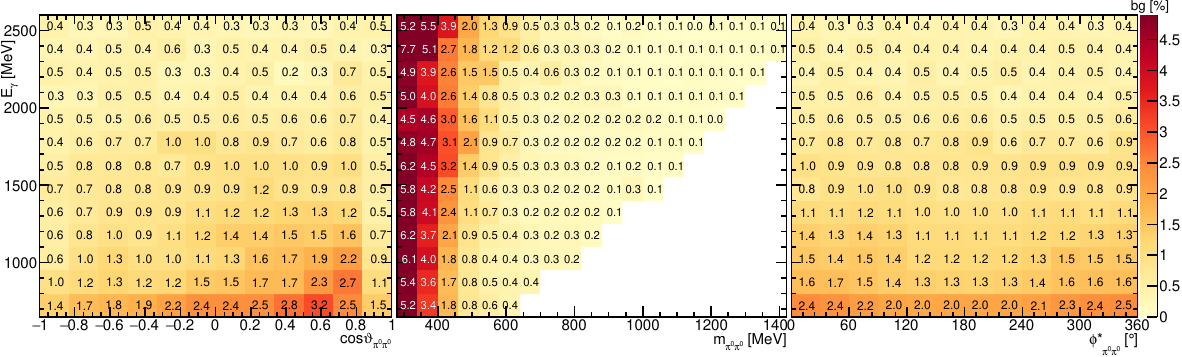}
} \caption{Fractional background contribution (in \%) depending on beam energy $E_{\Pgg}$ and $\cos\vartheta_{\Pgpz\Pgpz}$ (left),
$m_{\Pgpz\Pgpz}$ (center), or $\phi^*_{\Pgpz\Pgpz}$ (right), respectively.}
\label{fig:bg}
\end{figure*}

\begin{figure*}
\resizebox{\textwidth}{!}{%
  \includegraphics{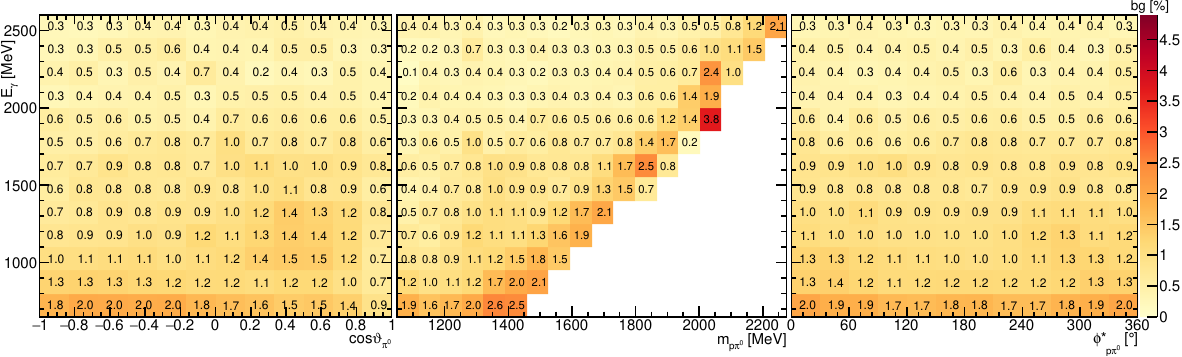}
} \caption{Fractional background contribution (in \%) depending on beam energy $E_{\Pgg}$ and $\cos\vartheta_{\Pgpz}$ (left),
$m_{\Pp\Pgpz}$ (center), or $\phi^*_{\Pp\Pgpz}$ (right), respectively.}
\label{fig:bg2}
\end{figure*}

\subsection{Background contribution}
\label{sec:bg}
It is not possible to estimate the (potentially polarized) background contamination of the selected events from the fit of $\Pgg\Pp\to\Pp_\text{miss}\Pgpz\Pgpz$ since signal and background cannot be separated in the fitted pion mass. Therefore, a less constrained kinematic fit with the hypothesis $\Pgg\Pp\to\Pp_\text{miss}\Pgpz\Pgg\Pgg$ was employed.
Here, the fitted pion is chosen as the best combination of all $\Pgg\Pgg$ pairs of the four photons. The remaining two photons are left unconstrained.

The determined amount of background $x$ will then not be used on an event-to-event basis in the maximum-likelihood fit (\textit{cf.} section~\ref{sec:lnL}) for extracting the observables. Instead, it will be used as a first order correction of the observables and as a source of the systematic uncertainty, \textit{cf.} section~\ref{sec:syst}.

The invariant mass of the unfitted (and uncut) photon pair in the hypothesis $\Pgg\Pp\to\Pp_\text{miss}\Pgpz\Pgg\Pgg$ showed a peak at the pion mass on a background distribution,
\textit{cf.} Fig.~\ref{fig:bg_fit}. The fit uses the best 2\Pgg{} combination for the fitted pion, which results in a dip on the pion peak.

The amount of background can vary significantly depending on the kinematic point. To allow for variations in the 5-dimensional phase space, the
background contribution was determined with methods of a multivariate side-band subtraction
\cite{williams}. This method, also known as the Q-factor method, assigns a quality factor to every
event which gives the probability for this event to originate from the signal sample. For a given
event, the nearest neighbors in the 5-dimensional phase space
were taken. The corresponding 2\Pgg-mass distribution was
then fitted with an unbinned maximum-likelihood fit by a sum of two normal distributions (peak with dip) and a polynomial background. For the fit to reliably converge the number of
nearest neighbors was increased until there were either 300 events in the background region ($m_{\Pgg\Pgg}<\unit[100]{MeV}$ or $>\unit[170]{MeV}$), 40 events in the lower and 65 in the higher mass background region, or 15000 nearest neighbors in total were reached. From the ratio of the background function to the function describing the signal the amount
of background at the position of the seed event (at the position of the vertical line in Fig.~\ref{fig:bg_fit}) was estimated. An example of such a fit is shown
in Fig.~\ref{fig:bg_fit}.

Originally, it was planned to provide the PWA group with the data on the polarization observables on an event-to-event basis and therefore also determine the background (and dilution factor, \textit{cf.} section~\ref{sec:dilution}) on an event-to-event basis. But since the data also contain beam asymmetry contributions from bound nucleons (with in principle unknown values of the beam asymmetries, \textit{cf.} section~\ref{sec:dilution}), at this time only binned data were provided and the averaged background within these bins ($x_j= {\langle1-Q_i\rangle}_{i\in\text{bin} j}$, where $x_j$ is the background in a one- or multi-dimensional bin) was used for the determination of the systematic uncertainty, \textit{cf.} section~\ref{sec:syst}. In addition, a correction of the observables has been performed as described below.

The background was determined to be at the order of \unit[1--2.5]{\%} for incident photon energies below
\unit[1500]{MeV} decreasing to below \unit[0.5]{\%} for higher energies, as shown in Fig.~\ref{fig:bg} and \ref{fig:bg2}. Only for the lowest
$m_{\Pgpz\Pgpz}$ significantly higher background of up to $\approx\unit[5]{\%}$ was found, see
Fig.~\ref{fig:bg}.

The observables shown in section~\ref{sec:results} were corrected by a factor $(1-x)^{-1}$, where $x$ denotes the amount of background, since the background events will lower
the value of the resulting polarization observable, assuming the background to be unpolarized. A possible polarization of the background events will be treated in the section on the systematic uncertainties (\textit{cf.} section~\ref{sec:syst}).

The background contribution was somewhat over-esti\-ma\-ted with the method described here. A test with simulated $\Pp4\Pgg$- and $\Pp\Pgpz\Pgpz$-final states, the former serving as the background contribution, revealed the determined background contribution to be about \unit[50]{\%} higher than expected from the simulated values. The effect of using the higher value is still covered by the systematic uncertainty $\Delta\text{BG}_\text{syst}$, which is (roughly) proportional to the determined background contribution (\textit{cf.} Eq.~\ref{eq:bgsyst}).

Correlations between the chosen nearest neighbors were found to be small:
The set of nearest neighbors of a typical event has an intersection only with about \unit[1]{\%} of all events and in this case the mean cardinality of the intersecting set is in the order of $<\unit[10]{\%}$ of the original sets cardinality.
Since the background and therefore also the systematic uncertainty on the observables were already somewhat overestimated, the correlation contributions were omitted in $\Delta\text{BG}_\text{syst}$.

\section{Data analysis}
\label{sec:analysis}
\begin{figure}
\resizebox{0.4\textwidth}{!}{%
  \includegraphics{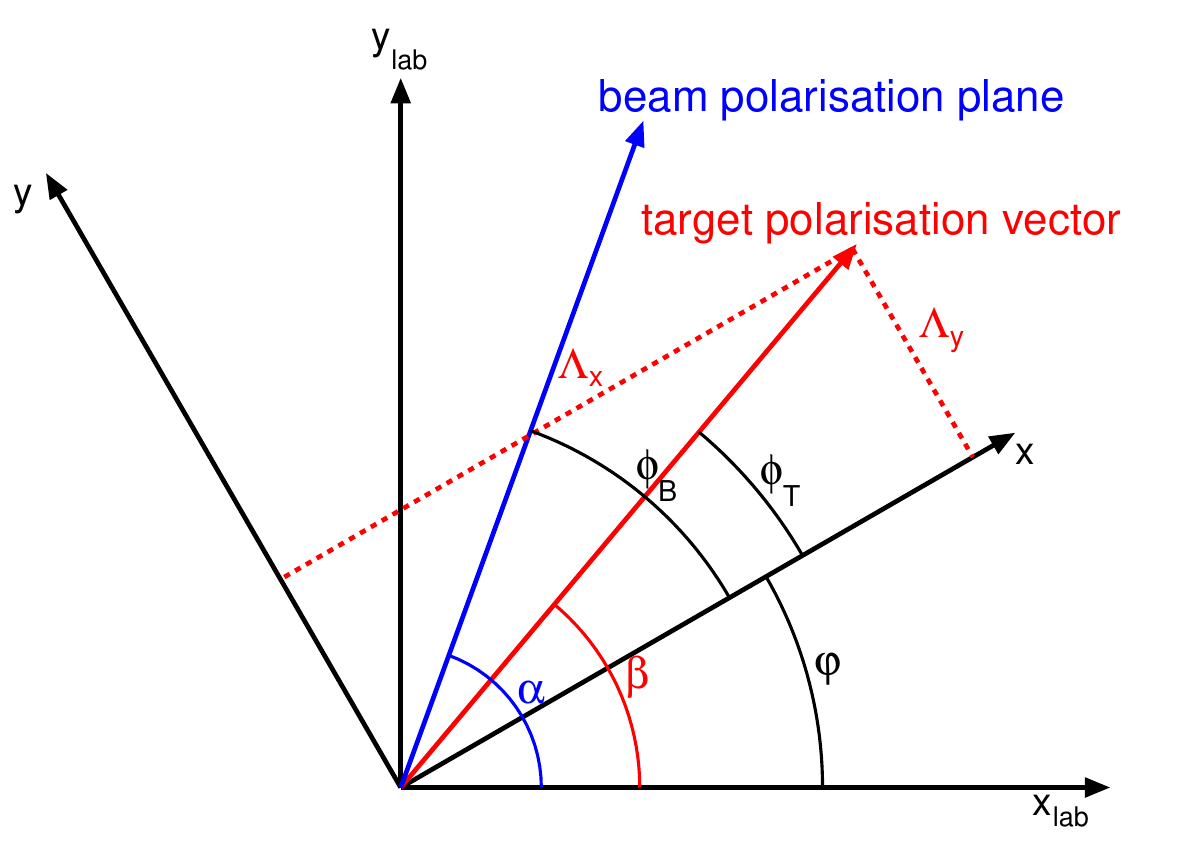}
}
\caption{Definition of the relevant angles in the laboratory frame and the reaction frame. Details see text.}
\label{fig:angles}
\end{figure}

The cross section of double pion photoproduction using a linearly polarized beam (polarization degree
$\delta_\ell$ with angle $\phi_B$ to the $x$-axis of the reaction plane) and a transversely polarized
target (polarization degree $\varLambda$ with angle $\phi_T$ to the $x$-axis of reaction plane) can
be written in the form (\textit{cf.} \cite{roberts}):
\begin{align}
	\dfrac{\dd\sigma}{\dd\varOmega} = \dfrac{\dd\sigma_0}{\dd\varOmega} & \cdot \Big\{1 +d\,\varLambda\cos(\beta-\varphi)\cdot {\rm P_x} +d\,\varLambda\sin(\beta-\varphi)\cdot {\rm P_y}\nonumber\\
	&+\delta_\ell \sin(2\alpha-2\varphi)\cdot {\rm I^s_\text{eff}}+\delta_\ell \cos(2\alpha-2\varphi)\cdot {\rm I^c_\text{eff}}\nonumber\\
	&+d\,\varLambda\,\delta_\ell\cos(\beta-\varphi)\sin(2\alpha-2\varphi)\cdot {\rm P_x^s} \nonumber\\
	&+d\,\varLambda\,\delta_\ell\sin(\beta-\varphi)\sin(2\alpha-2\varphi)\cdot {\rm P_y^s}\nonumber\\
	&+d\,\varLambda\,\delta_\ell\cos(\beta-\varphi)\cos(2\alpha-2\varphi)\cdot {\rm P_x^c} \nonumber\\
	&+d\,\varLambda\,\delta_\ell\sin(\beta-\varphi)\cos(2\alpha-2\varphi)\cdot {\rm P_y^c} \Big\} ,
\label{eq:dsigpol}
\end{align}
where $\frac{\dd\sigma_0}{\dd\varOmega}$ denotes the unpolarized cross section, $\rm P_x$ and $\rm P_y$ are
target asymmetries, $\rm I^s_\text{eff}$ and $\rm I^c_\text{eff}$ beam asymmetries, and ${\rm P_x^s}, {\rm P_y^s},
{\rm P_x^c}$ and $\rm P_y^c$ double polarization observables. The photon polarization plane and
the target polarization vector are oriented with respective angles $\alpha$ and $\beta$ to the $x$-axis in the laboratory frame.
The reaction plane is rotated by an angle $\varphi$ relative to the laboratory frame, as depicted
in Fig.~\ref{fig:angles}. The observables were determined using a butanol target which contained
polarizable free protons as well as unpolarized bound nucleons in the carbon and oxygen nuclei, \textit{i.e.} ${\rm I^c_\text{eff}}=d{\rm I^c}+(1-d){\rm I^c_\text{bound}}$.
Thus, the target polarization was effectively reduced by the so-called dilution factor $d$ which is
the fraction of polarizable free protons. The measured beam asymmetries $\rm I^s_\text{eff}$ and
$\rm I^c_\text{eff}$ are a mixture of the beam asymmetries from free protons and those from the bound
nucleons.

If the kinematic variable $\phi^*$ is integrated out, half of the polarization observables in Eq.~(\ref{eq:dsigpol}) vanish, namely $\rm P_x$,
$\rm I^s_\text{eff}$, $\rm P_y^s$, and $\rm P_x^c$. When integrating over $\phi^*$ and $\theta^*$, the kinematic
mimics the situation of single-meson photoproduction and the remaining observables are often identified
with the single-meson polarization observables \cite{barker}:
\begin{align}
	\dfrac{\dd\sigma}{\dd\varOmega} = \dfrac{\dd\sigma_0}{\dd\varOmega} & \cdot\,\Big\{1+d\,\varLambda\,{\rm T}\,\sin(\beta-\varphi)\nonumber\\
	&-\delta_\ell\,\varSigma_\text{eff}\cos(2\alpha-2\varphi) \nonumber\\
	&-d\,\varLambda\,\delta_\ell\,{\rm H}\,\sin(2\alpha-2\varphi)\,\cos(\beta-\varphi) \nonumber\\
	&-d\,\varLambda\,\delta_\ell\,{\rm\tilde P}\,\cos(2\alpha-2\varphi)\,\sin(\beta-\varphi) \Big\} ,
\end{align}
where $\rm P_y$ becomes $\rm T$, ${\rm I^c_\text{eff}}=-\varSigma_\text{eff}$, ${\rm P_x^s}=-{\rm H}$, and ${\rm P_y^c}=-{\rm \tilde P}$.\footnote{In single-meson
photoproduction the double polarization observable denoted as $\rm \tilde P$ here is identical to the recoil
asymmetry, usually denoted as $\rm P$. In double meson photoproduction this is in general not the case.}

\subsection{Dilution factor}
\label{sec:dilution}
\begin{figure}
\resizebox{0.5\textwidth}{!}{%
  \includegraphics{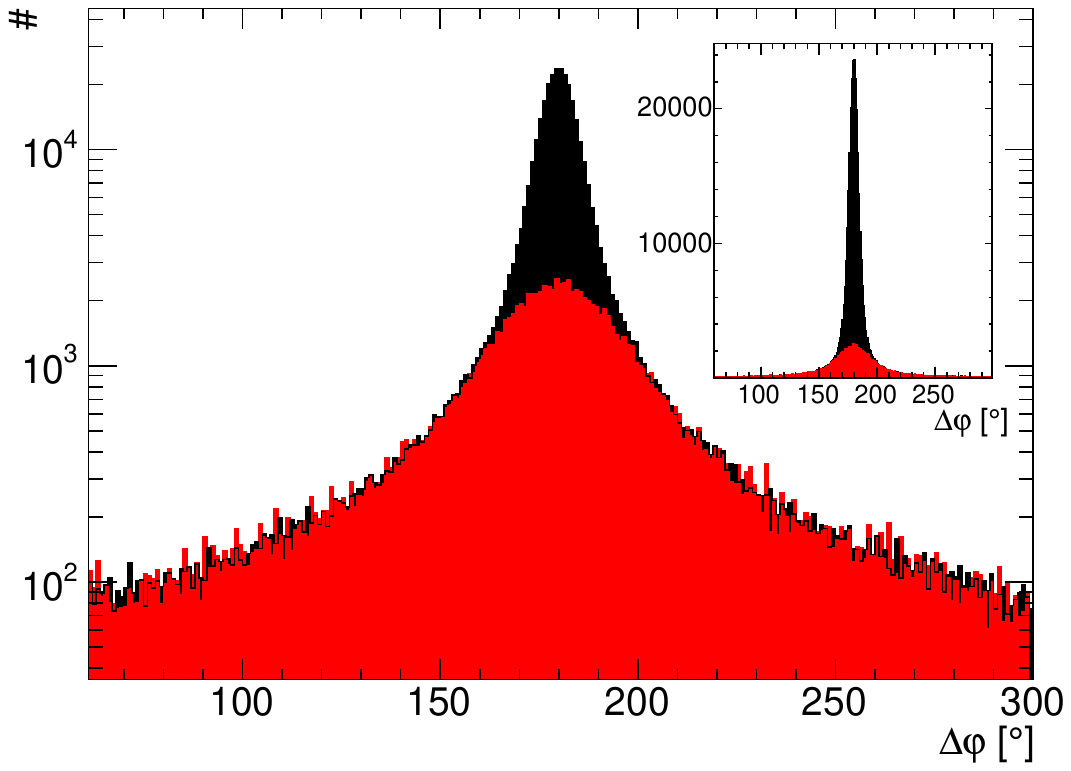}
} \caption{Coplanarity spectrum from data taken with the butanol target (black) and with the carbon
target (red). The latter already scaled. Events with $\unit[650]{MeV} < E_{\Pgg} < \unit[2600]{MeV}$ are shown. The insert shows the same spectrum on a linear axis.}
\label{fig:copl}
\end{figure}

The contributions from the bound nucleons were determined by placing the carbon foam target of density
close to that of the carbon and oxygen nuclei in the butanol target within the cryostat.
The azimuthal angle difference between measured proton and the four photons (coplanarity) was used
to calculate the dilution factor. Since the bound nuclei are subject to Fermi motion their
coplanarity spectrum is broadened compared to the free protons, as shown in Fig.~\ref{fig:copl}.

\begin{figure}
\resizebox{0.475\textwidth}{!}{%
  \includegraphics{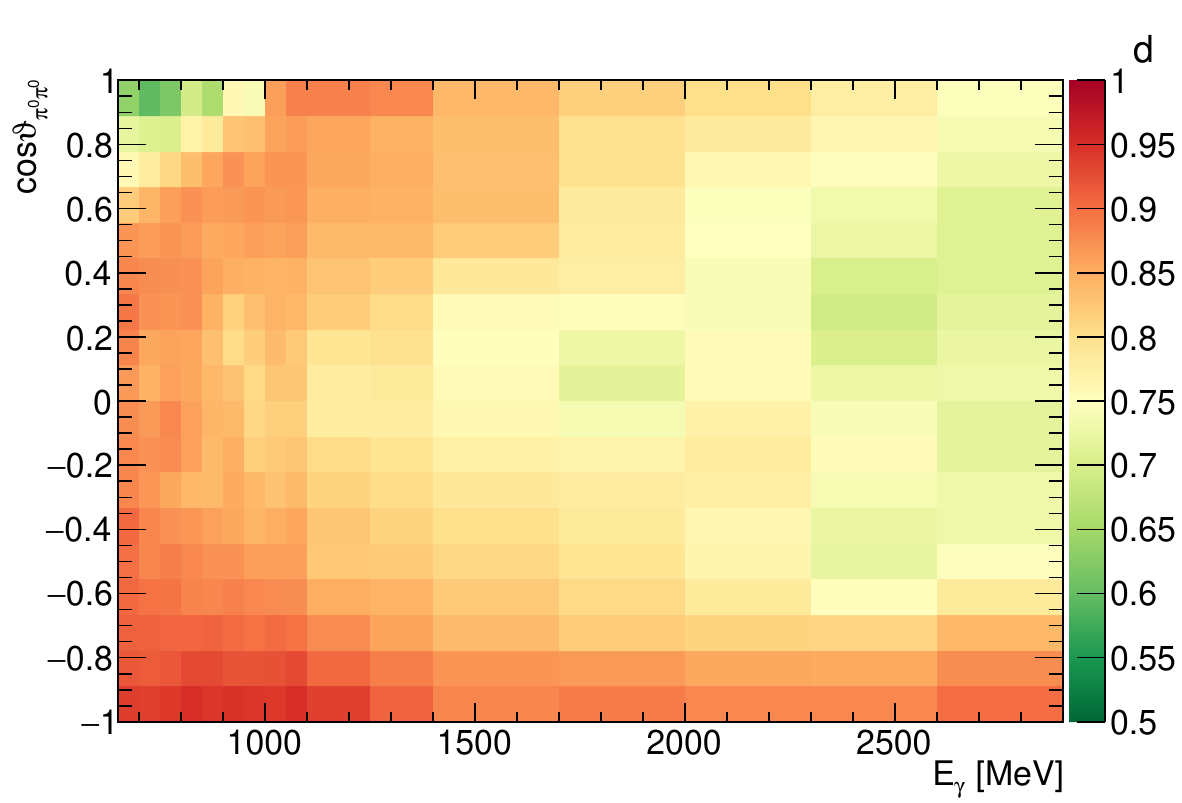}
} \caption{Value of the dilution factor depending on $E_{\Pgg}$ and
$\cos\vartheta_{\Pgpz\Pgpz}$.}
\label{fig:dil}
\end{figure}

Differences in the photon flux ($\Phi$) of butanol and carbon data were accounted for
by an energy-dependent weighting factor which is the ratio of the measured photon fluxes: $w=\Phi_\text{But}/\Phi_\text{C}$. Additional small
differences in the target area density were treated by a global scaling factor $s=1.008\pm0.009$,
which was determined from the ratio of the butanol to the carbon events\footnote{Selected as described in section~\ref{sec:selection} but without the cut on the coplanarity.} in the outer parts of the coplanarity spectrum (below $155^\circ$ or above $205^\circ$, corresponding to $\approx5\sigma$, \textit{cf.} Fig.~\ref{fig:copl}).

Using the global scaling factor $s$, it was then possible to determine the dilution factor on an event-to-event basis using the
nearest neighbor method similar to what is described in section~\ref{sec:bg}. Here, a number $n_\text{B}$ of nearest neighbors
from the butanol data were taken which defines a hyper-sphere in the 5-dimensional phase space.
Then the number $n_\text{C}$ of events from the carbon target within the same hyper-sphere were
determined. The dilution factor is then calculated as
\begin{equation*}
 d = \dfrac{n_\text{B} - w\cdot s\cdot n_\text{C}}{n_\text{B}}.
\end{equation*}
To keep the volume of the hyper-sphere (and thus the correlations between events) as small as possible the size was chosen such that 10 events
(before weighting) from the carbon data were inside.

To show the variation of the dilution factor with the kinematic variables, the event-based dilution
factors were sorted into bins. Fig.~\ref{fig:dil} shows an example of such a histogram binned in
$E_{\Pgg}$ and $\cos\vartheta_{\Pgpz\Pgpz}$.

The event-based results are consistent with a previously performed 2-dimensional binned analysis, which did not extend to the highest $E_{\Pgg}$ presented here.

\subsection{Event-based maximum-likelihood fit}
\label{sec:lnL}
The polarization observables were determined by an event-based maximum-likelihood fit to the
azimuthal distribution of the events. Therefore, it was not necessary to bin the data in the
azimuthal angle $\varphi$ but only in the kinematic variables (\textit{cf.} section~\ref{sec:results}). The azimuthal variation of the
polarized cross section was expanded in a Fourier series\footnote{A global normalization factor of
$(2\pi)^{-1}$ only leads to a global shift in the log-likelihood and can therefore be omitted.}
\begin{equation}
 f_\text{phy} = \dfrac{\dd\sigma/\dd\varOmega}{\dd\sigma_0/\dd\varOmega} = 1+
 \sum\limits_{k=1}^3 [a_k \sin(k\,\varphi) + b_k \cos(k\,\varphi)],
\end{equation}
where the coefficients $a_k$ and $b_k$ were determined from Eq.~(\ref{eq:dsigpol}) and include only
the eight polarization observables as free parameters. They are given by:
\begin{align*}
	a_1 =& +d\varLambda {\rm P_x} \sin(\beta) -d\varLambda {\rm P_y} \cos(\beta) \\
	& -\frac{1}{2}\delta_\ell d\varLambda [({\rm P_x^s}-{\rm P_y^c}) \left( \cos(2\alpha)\cos(\beta) + \sin(2\alpha)\sin(\beta) \right) \\
	& -({\rm P_x^c}+{\rm P_y^s}) \left( \sin(2\alpha)\cos(\beta) - \cos(2\alpha)\sin(\beta) \right) ] \\
	b_1 =& +d\varLambda {\rm P_x} \cos(\beta) +d\varLambda {\rm P_y} \sin(\beta) \\
	& +\frac{1}{2}\delta_\ell d\varLambda [({\rm P_x^s}-{\rm P_y^c}) \left( \sin(2\alpha)\cos(\beta) - \cos(2\alpha)\sin(\beta) \right) \\
	& +({\rm P_x^c}+{\rm P_y^s}) \left( \cos(2\alpha)\cos(\beta) + \sin(2\alpha)\sin(\beta) \right) ] \\
	a_2 =& -\delta_\ell {\rm I^s_\text{eff}}\cos(2\alpha) +\delta_\ell {\rm I^c_\text{eff}}\sin(2\alpha) &\\
	b_2 =& +\delta_\ell {\rm I^s_\text{eff}}\sin(2\alpha) +\delta_\ell {\rm I^c_\text{eff}}\cos(2\alpha) &\\
	a_3 =& -\frac{1}{2}\delta_\ell d\varLambda [({\rm P_x^s}+{\rm P_y^c}) \left( \cos(2\alpha)\cos(\beta) - \sin(2\alpha)\sin(\beta) \right) \\
	& -({\rm P_x^c}-{\rm P_y^s}) \left( \sin(2\alpha)\cos(\beta) + \cos(2\alpha)\sin(\beta) \right) ] \\
	b_3 =& +\frac{1}{2}\delta_\ell d\varLambda [({\rm P_x^s}+{\rm P_y^c}) \left( \sin(2\alpha)\cos(\beta) + \cos(2\alpha)\sin(\beta) \right) \\
	& +({\rm P_x^c}-{\rm P_y^s}) \left( \cos(2\alpha)\cos(\beta) - \sin(2\alpha)\sin(\beta) \right) ] \ .
\end{align*}

Additionally, potential detector asymmetries were expanded in a Fourier series as well:
\begin{equation}
 f_\text{det} = 1+ \sum\limits_{k=1}^\infty [c_k \sin(k\,\varphi) + d_k \cos(k\,\varphi)].
\end{equation}
Those two series had to be multiplied and normalized, leading to
\begin{equation}
 f_\text{sig} = \dfrac{f_\text{phy}\cdot f_\text{det}}{
 \int f_\text{phy} f_\text{det}\ \dd\varphi} = \dfrac{f_\text{phy}\cdot f_\text{det}}{1+\frac{1}{2}\sum_{k=1}^3 (a_k c_k + b_k d_k)}.
\label{eq:fsig}
\end{equation}
Since the series representing the physical signal $f_\text{phy}$ stops at $k=3$, the coefficients can only be influenced
from coefficients of the detector series $f_\text{det}$ up to twice that number\footnote{This
follows directly from the product-to-sum identities, \textit{e.g.} $\cos x\cos y =
\nicefrac{1}{2}(\cos(x-y)+\cos(x+y))$.}. Therefore, the series expansion of $f_\text{det}$ could be
stopped at $k=6$.
The detector efficiency coefficients ($c,d$) decouple from the physics coefficients ($a,b$) by changing the polarization directions: $\varLambda\mapsto-\varLambda$ or $\delta_\ell\mapsto-\delta_\ell$.

To account for random time background in the data, the likelihood function $\mathcal L$ also
contained a function of form Eq.~\ref{eq:fsig}, called $f_\text{tbg}$ for those events. The
coefficients of this function are determined with events in a side-band region. Thus, the following
expression was minimized:
\begin{align}
 -\ln\mathcal L = &-\sum\limits_{i=1}^{N_\text{sig}}\ln\left( \xi\,f_\text{sig}(\varphi_i) + (1-\xi)\,f_\text{tbg}(\varphi_i) \right) \nonumber\\
 & -\sum\limits_{j=1}^{N_\text{tbg}}\ln\left( f_\text{tbg}(\varphi_j) \right),
\end{align}
where $\xi=\frac{N_\text{sig}-R\cdot N_\text{tbg}}{N_\text{sig}}$ and $R=\nicefrac{1}{20}$ is the ratio of the time cut width to the side-band width.

Together with the eight observables\footnote{Different polarization settings are already needed, in order to extract the eight observables from the six coefficients $a,b$. This effectively disentangles the double polarization observables from the target polarization observables in $a_1$ and $b_1$.} and twelve detector asymmetry coefficients in the signal part,
there are the same number of parameters in the time background part, leading to a total of 40 free parameters.

\subsection{Systematic uncertainty}
\label{sec:syst}
The main sources of the systematic uncertainty are the uncertainty of the beam and target
polarization values, the uncertainty of the dilution factor, and the amount of background events in
the data sample. The relative uncertainties of the first three contributions are directly
proportional to the absolute value of the polarization observable. In order not to under-estimate
the systematic uncertainty for very small observables\footnote{For example, an observable of absolute value $0.0\pm0.1$ would have a systematic uncertainty from the polarization uncertainties of 0.} the relative uncertainties of the first three contributions are multiplied with a convolution of the absolute value of the corrected (see below) measured observable $\mathcal O/(1-x)$ with a
Gaussian which has the statistical uncertainty $\sigma_s$ of the observable as its width:
\begin{equation}
 \mathcal{O}^\prime := \int\limits_{-5\sigma_s}^{5\sigma_s} \left| \frac{\mathcal{O}}{1-x} - \omega\right| \cdot \dfrac{1}{\sqrt{2\pi\sigma_s^2}}\cdot \exp\left(-\frac{\omega^2}{2\sigma_s^2}\right)\,\dd\omega .
\end{equation}
In the numerical integration, a precision of $10^{-6}$ can be achieved with integration limits of $\pm5\sigma_s$.

The systematic uncertainty due to the amount of background $\Delta\text{BG}_\text{syst}$ is
independent of the value of the polarization observable. The true observable $\mathcal O_\text{S}$
is related to the measured one ($\mathcal O$) and the possible background observable $\mathcal
O_\text{BG}$ by
\begin{equation}
 \mathcal O=(1-x)\cdot\mathcal O_\text{S} + x\cdot\mathcal O_\text{BG} ,
\end{equation}
where $x$ is the amount of background. Dividing the measured observable by $(1-x)$ is a first order
correction but a possible contribution of polarized background remains.

The uncertainty due to the background can be estimated by the expected value of the quadratic difference of the
corrected measured observable $\frac{\mathcal O}{1-x}$ to the true value $\mathcal O_\text{S}$:
\begin{align}
 (\Delta\text{BG}_\text{syst})^2 &= \operatorname{E}\left[\left(\frac{\mathcal O}{1-x}-\mathcal O_\text{S}\right)^2\right] .
\end{align}
Since nothing is known about the background observable, a uniform distribution between $-1$ and
$1$ was assumed (principle of maximum entropy), leading to the following expression for the
systematic uncertainty:
\begin{equation}
 (\Delta\text{BG}_\text{syst})^2 =\frac{1}{2}\int\limits_{-1}^{1} \left(\frac{x}{1-x}\right)^2
 \mathcal O_\text{BG}^2 \dd\mathcal O_\text{BG} = \left(\frac{x}{1-x}\right)^2 \cdot\frac{1}{3}.
 \label{eq:bgsyst}
\end{equation}

\begin{figure}
\resizebox{.5\textwidth}{!}{%
  \includegraphics{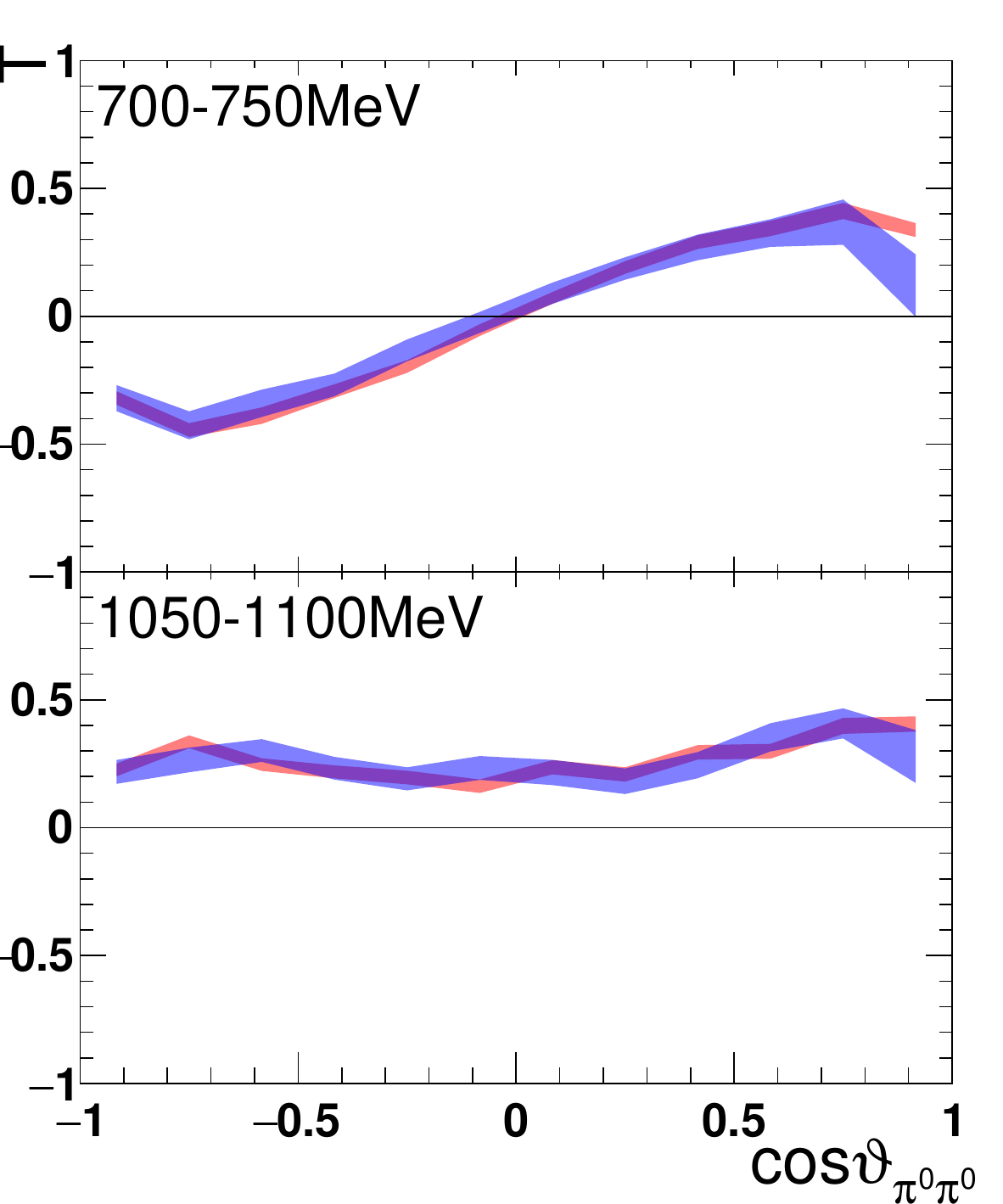}
  \includegraphics{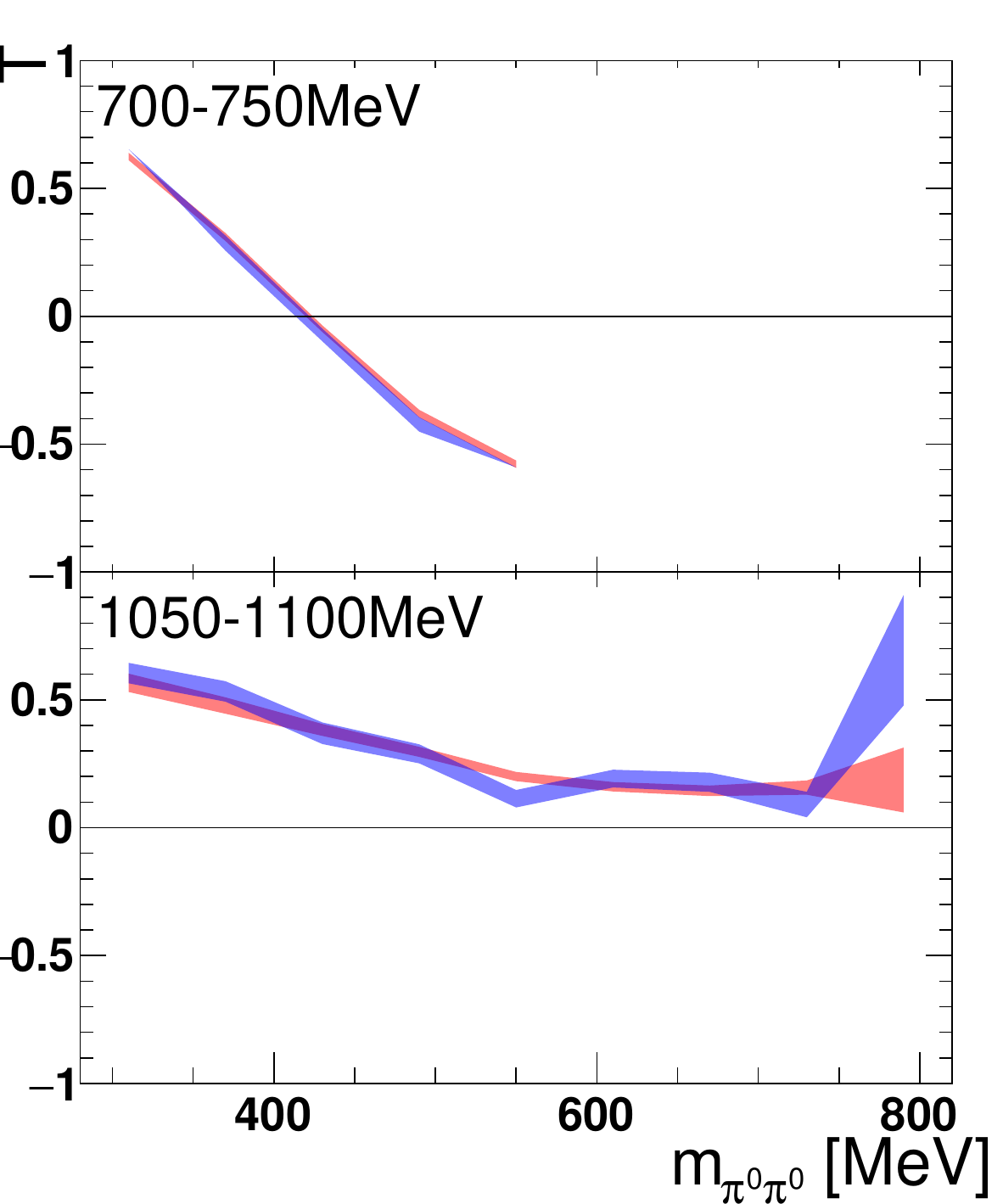}
} \caption{Example of the predicted target asymmetry $\rm T$ for two beam energies $E_{\Pgg}$ as a
function of $\cos\vartheta_{\Pgpz\Pgpz}$ or $m_{\Pgpz\Pgpz}$ determined from generated Monte Carlo
events (red band). The blue band results from taking the acceptance into account. The widths of the bands are given by the statistical uncertainties of the MC event sample.}
\label{fig:T_PWA}
\end{figure}

Systematic effects due to limited acceptance in the 5D phase space were investigated by Monte Carlo
events with polarization weights from a BnGa PWA solution. The observables from generated events
(with perfect acceptance) were compared to the reconstructed events (taking the detector acceptance
into account). Examples of such comparisons are shown in Fig.~\ref{fig:T_PWA}. The mean of the
distribution of differences mostly vanished within $1\sigma$ of the statistical uncertainty of the difference of the MC event sample. Therefore, no significant deviation due to
the acceptance was found.

The individual systematic uncertainties are added qua\-dratically, leading to
\begin{align}
 \Delta\mathcal O_\text{syst}^2 = &\left[\left(\frac{\Delta\delta_{\ell,
 \text{syst}}}{\delta_\ell}\right)^2 + \left(\frac{\Delta\varLambda_\text{syst}}{\varLambda}\right)^2 +
 \left(\frac{\Delta d_\text{syst}}{d}\right)^2\right]\mathcal{O}^{\prime\,2} \nonumber \\
 &+ (\Delta\text{BG}_\text{syst})^2 .
\end{align}
The systematic uncertainty of the dilution factor can be related to the uncertainty of the scaling factor $s$ by
\begin{equation}
 \left(\frac{\Delta d_\text{syst}}{d}\right)^2 = \left(\frac{\Delta s_\text{syst}}{s} \cdot \frac{1-d}{d}\right)^2 .
\end{equation}
The systematic uncertainty of the scaling factor was determined from the spread of an energy-dependent determination of the scaling factor, which resulted in $\Delta s_\text{syst}/s = \unit[5]{\%}$.
The uncertainty from the ratio of the photon fluxes $w$ is negligible in the dilution factor uncertainty because of the high statistics of the flux measurements.

For the target asymmetries no beam polarization is necessary. Hence, the systematic uncertainty of
the beam polarization is not added in that case. Typical values for the relative polarization
uncertainties are \unit[5]{\%} (beam) and \unit[2]{\%} (target), for the relative dilution factor
uncertainty $\lesssim\unit[1]{\%}$, and for $\Delta\text{BG}_\text{syst}$ mostly about 0.01 with only a few areas exceeding 0.03 (\textit{cf.} section~\ref{sec:bg}).

In most cases, the total systematic uncertainty amounts to $\lesssim 0.03$.

\begin{figure*}
\begin{center}
\resizebox{.85\textwidth}{!}{%
  \includegraphics{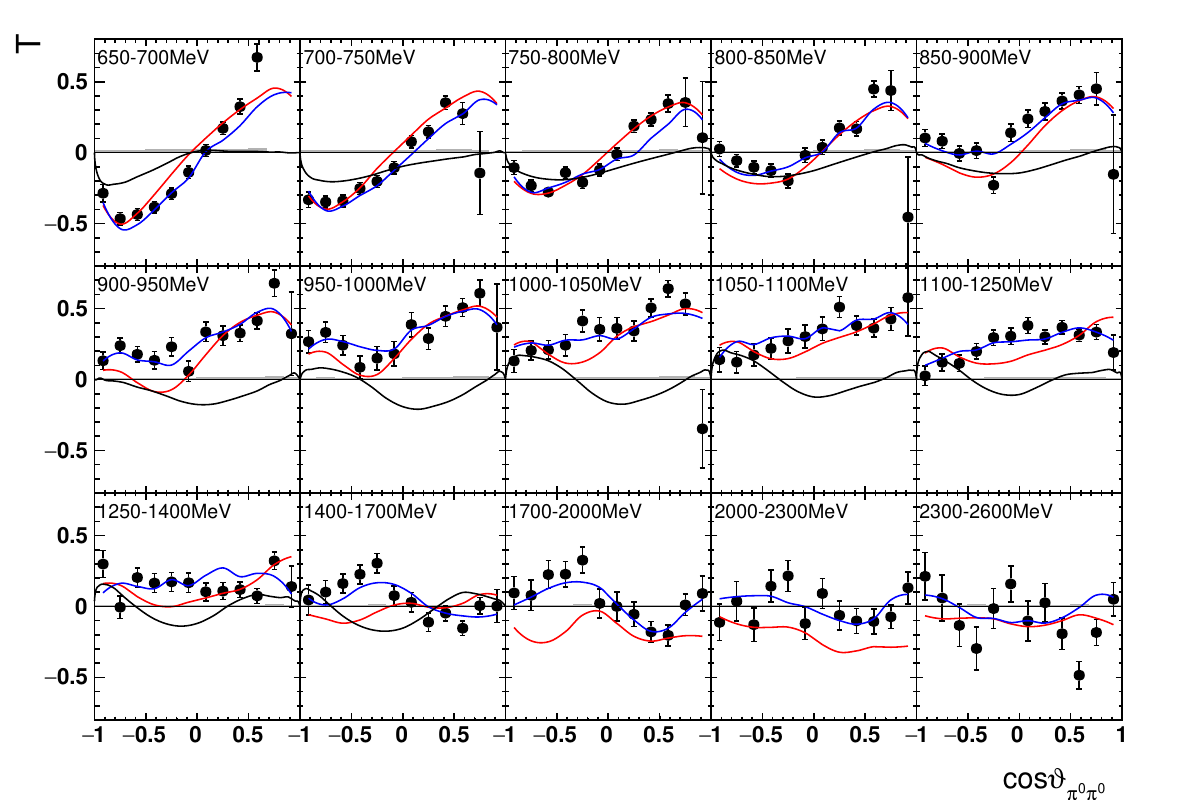}
} \caption{The target asymmetry $\rm T$ as a function of beam energy $E_{\Pgg}$ and
$\cos\vartheta_{\Pgpz\Pgpz}$. The colored lines represent PWA solutions: $2\pi$-MAID in black, BnGa 2014-02 in red, new BnGa 2022-02 in blue. The systematic uncertainty is shown as a gray band.}
\label{fig:T_costh}
\end{center}
\begin{center}
\resizebox{.85\textwidth}{!}{%
  \includegraphics{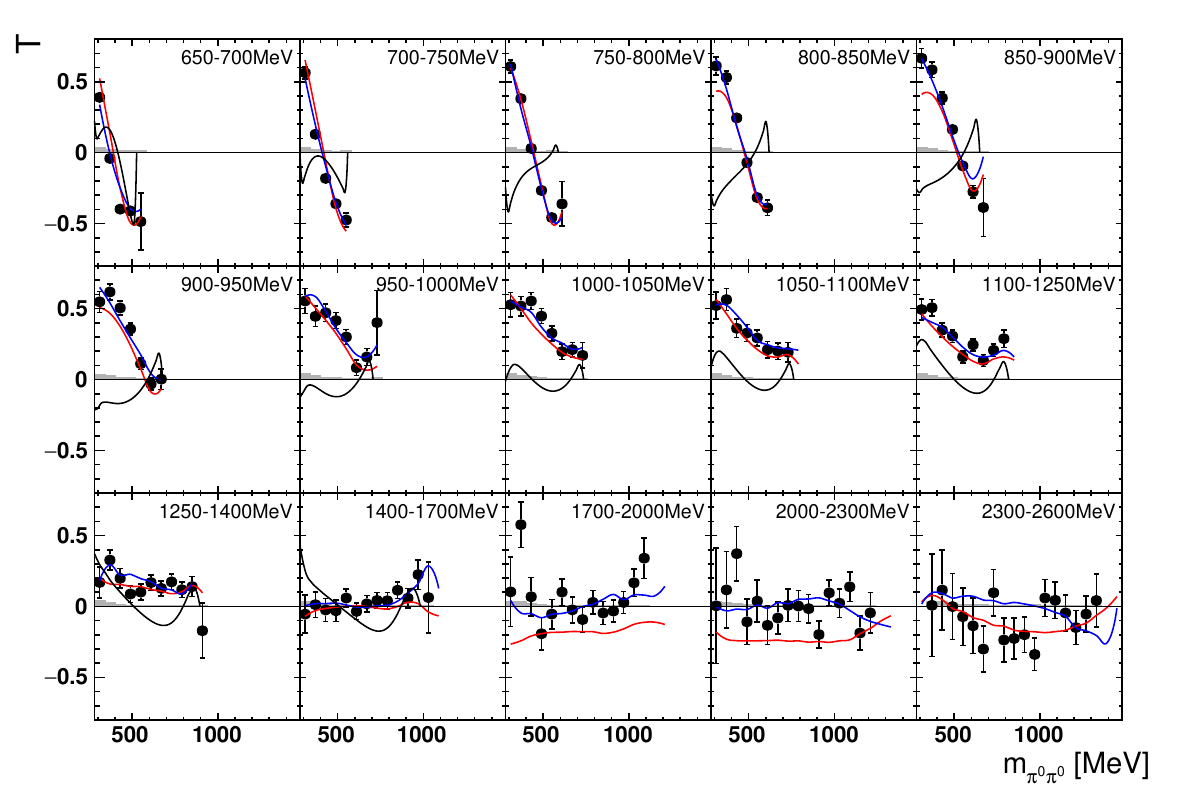}
} \caption{The target asymmetry $\rm T$ as a function of beam energy $E_{\Pgg}$ and $m_{\Pgpz\Pgpz}$. The colored lines represent PWA solutions: $2\pi$-MAID in black, BnGa 2014-02 in red, new BnGa 2022-02 in blue. The systematic uncertainty is shown as a gray band.}
\label{fig:T_IM}
\end{center}
\end{figure*}

\begin{figure*}
\begin{center}
\resizebox{.82\textwidth}{!}{%
  \includegraphics{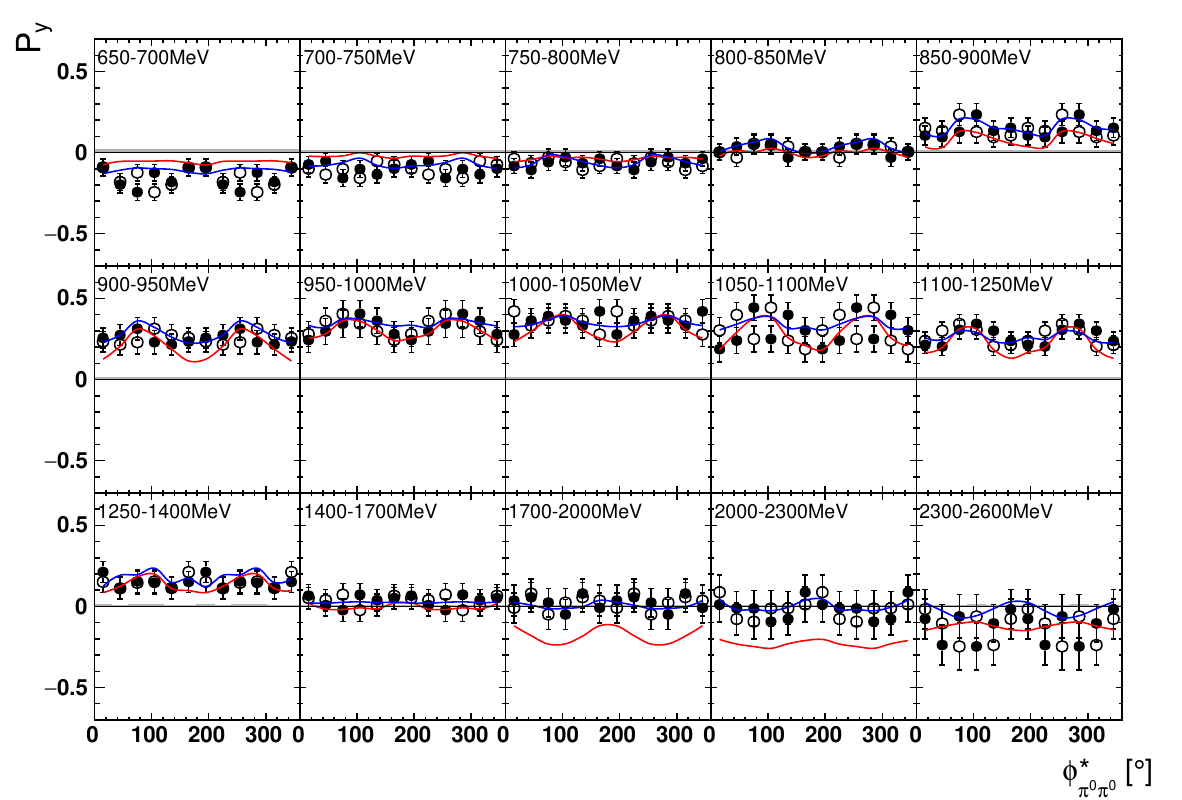}
} \caption{The target asymmetry $\rm P_y$ as a function of beam energy $E_{\Pgg}$ and $\phi^*_{\Pgpz\Pgpz}$. The
open symbols make use of the symmetry properties. The colored lines represent PWA solutions: BnGa 2014-02 in red, new BnGa 2022-02 in blue. The systematic uncertainty is shown as a gray band.}
\label{fig:Py}
\end{center}
\begin{center}
\resizebox{.82\textwidth}{!}{%
  \includegraphics{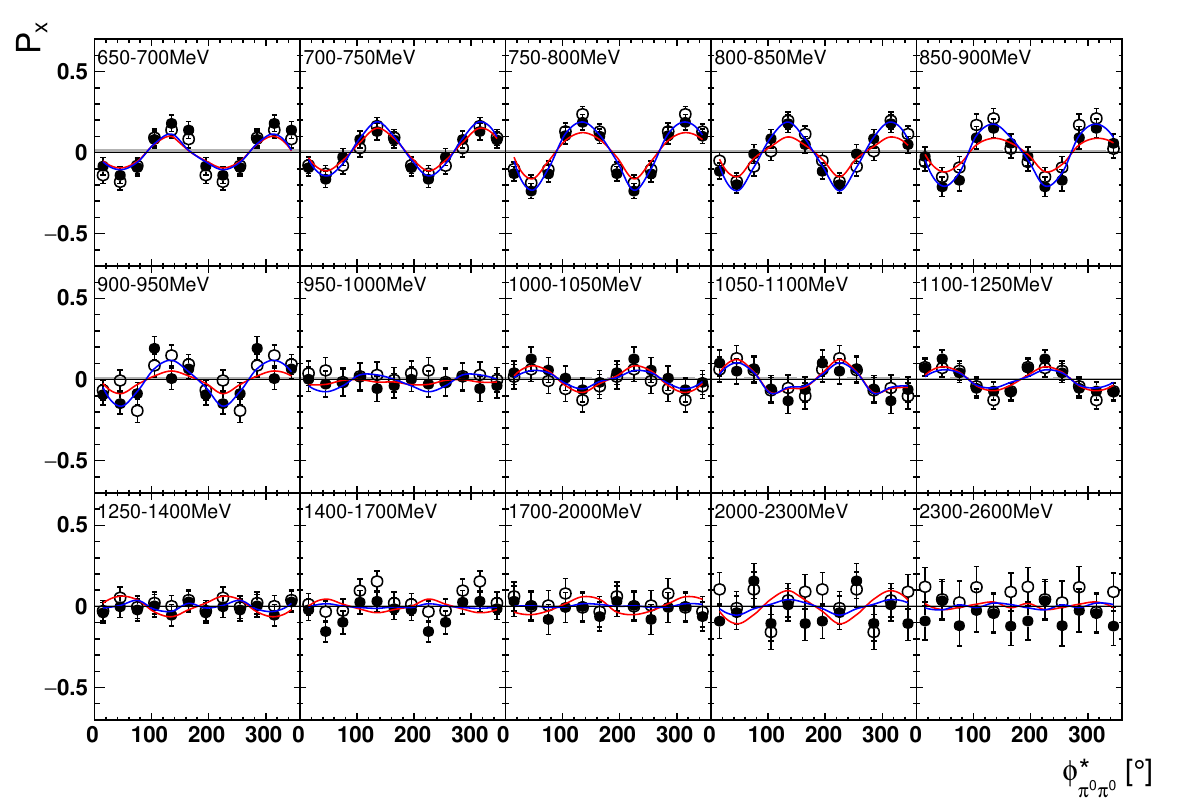}
} \caption{The target asymmetry $\rm P_x$ as a function of beam energy $E_{\Pgg}$ and $\phi^*_{\Pgpz\Pgpz}$. The
open symbols make use of the symmetry properties. The colored lines represent PWA solutions: BnGa 2014-02 in red, new BnGa 2022-02 in blue. The systematic uncertainty is shown as a gray band.}
\label{fig:Px}
\end{center}
\end{figure*}

\begin{figure*}
\begin{center}
\resizebox{.85\textwidth}{!}{%
  \includegraphics{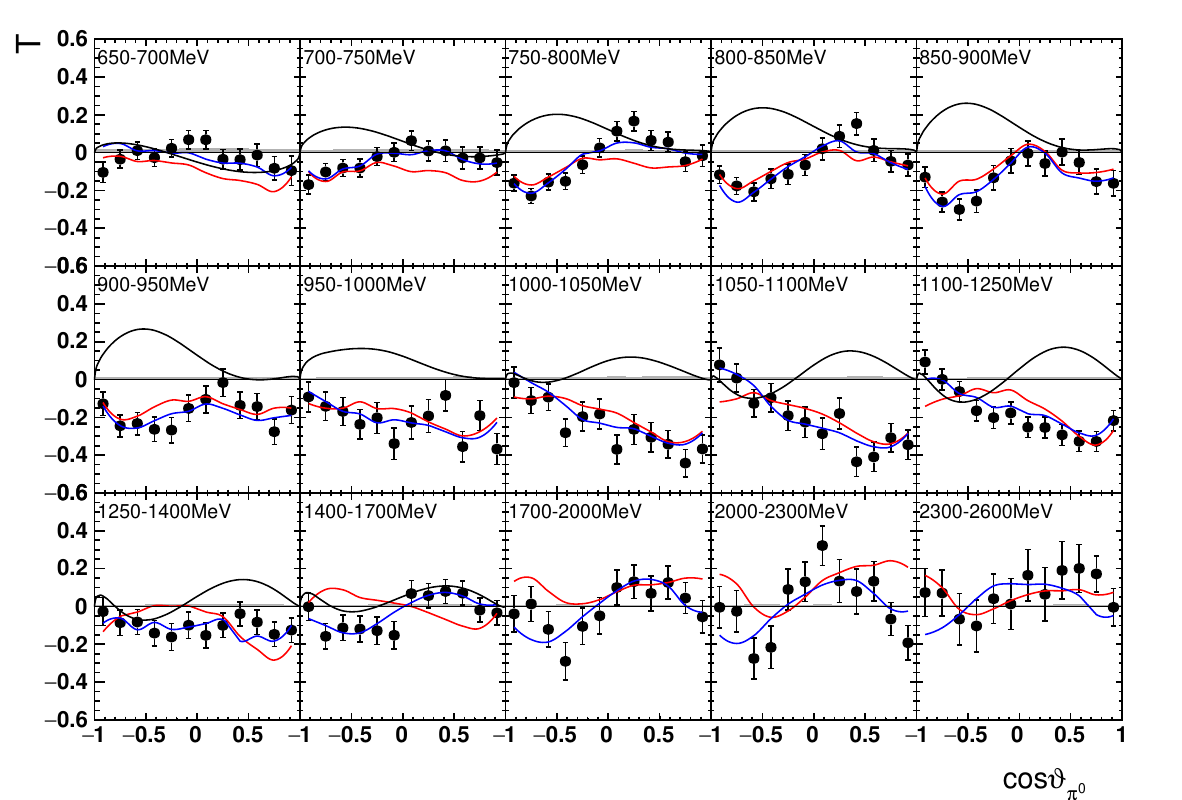}
} \caption{The target asymmetry $\rm T$ as a function of beam energy $E_{\Pgg}$ and
$\cos\vartheta_{\Pgpz}$. The colored lines represent PWA solutions: $2\pi$-MAID in black, BnGa 2014-02 in red, new BnGa 2022-02 in blue. The systematic uncertainty is shown as a gray band.}
\label{fig:T_costh_ppi0}
\end{center}
\begin{center}
\resizebox{.85\textwidth}{!}{%
  \includegraphics{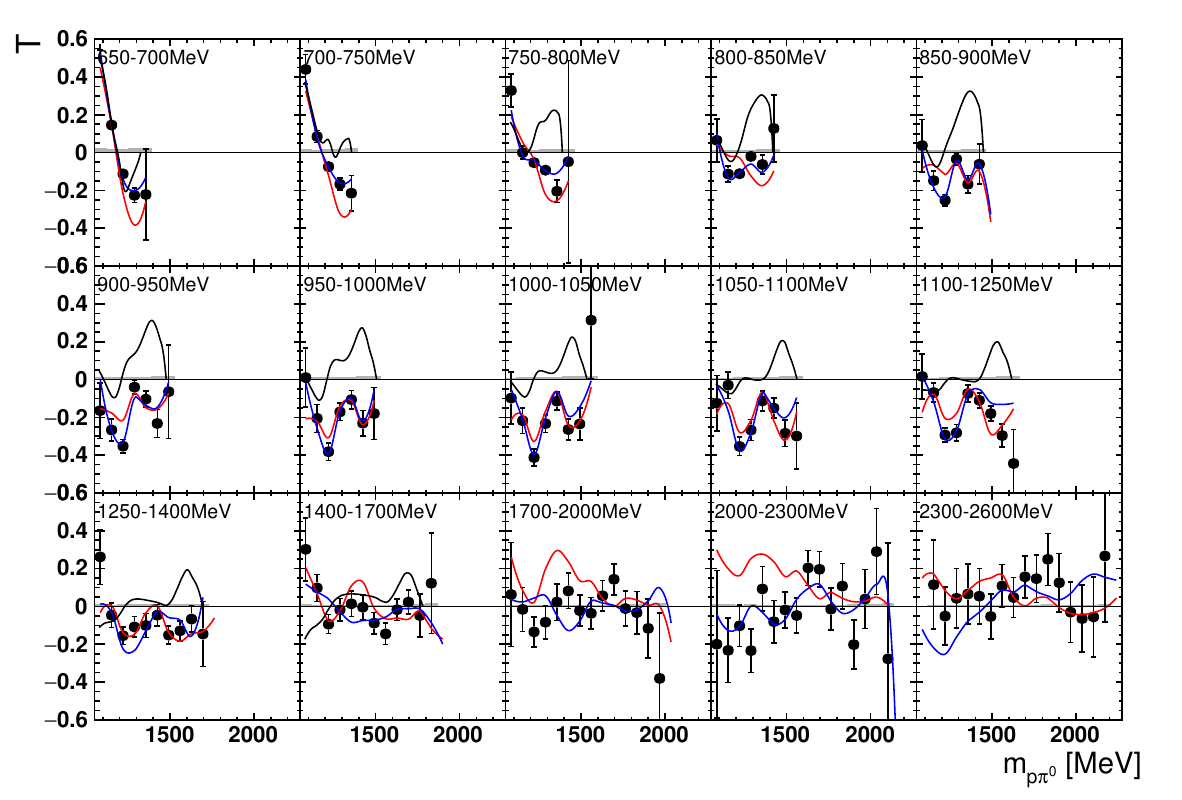}
} \caption{The target asymmetry $\rm T$ as a function of beam energy $E_{\Pgg}$ and $m_{\Pp\Pgpz}$. The colored lines represent PWA solutions: $2\pi$-MAID in black, BnGa 2014-02 in red, new BnGa 2022-02 in blue. The systematic uncertainty is shown as a gray band.}
\label{fig:T_IM_ppi0}
\end{center}
\end{figure*}

\begin{figure*}
\begin{center}
\resizebox{.82\textwidth}{!}{%
  \includegraphics{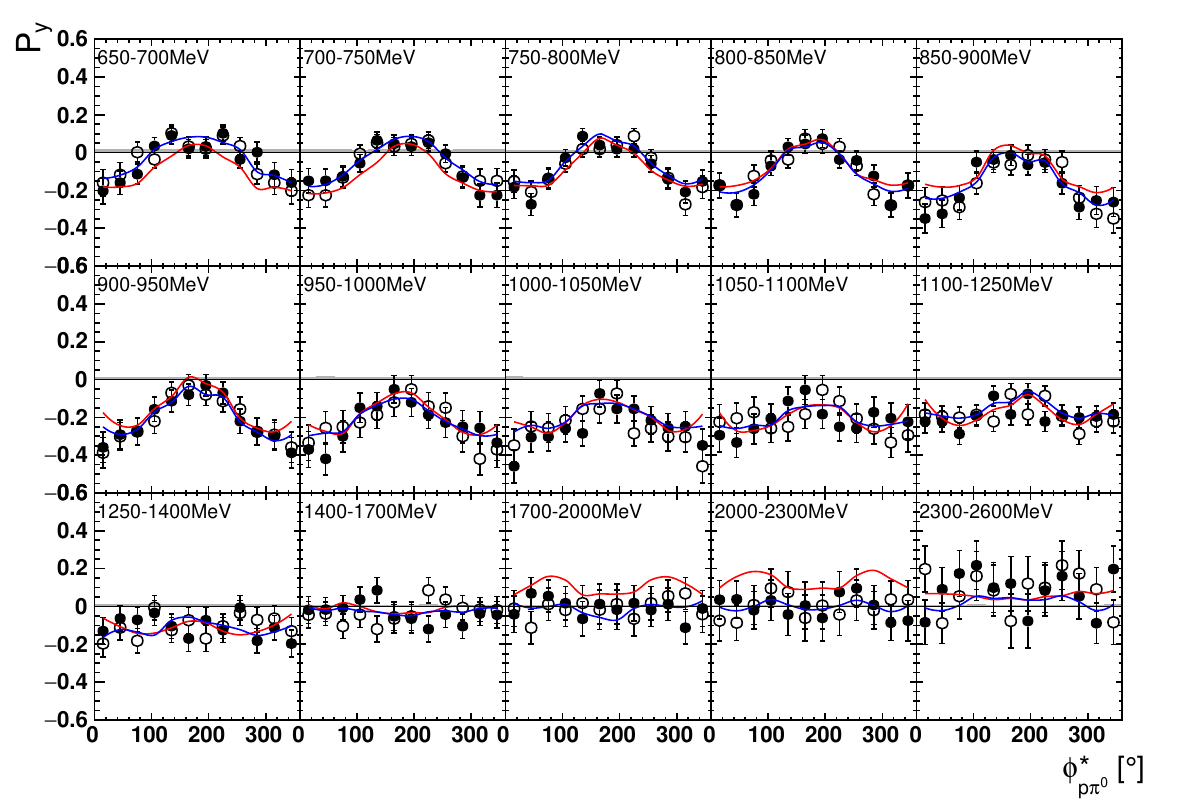}
} \caption{The target asymmetry $\rm P_y$ as a function of beam energy $E_{\Pgg}$ and $\phi^*_{\Pp\Pgpz}$. The
open symbols make use of the symmetry properties. The colored lines represent PWA solutions: BnGa 2014-02 in red, new BnGa 2022-02 in blue. The systematic uncertainty is shown as a gray band.}
\label{fig:Py_ppi0}
\end{center}
\begin{center}
\resizebox{.82\textwidth}{!}{%
  \includegraphics{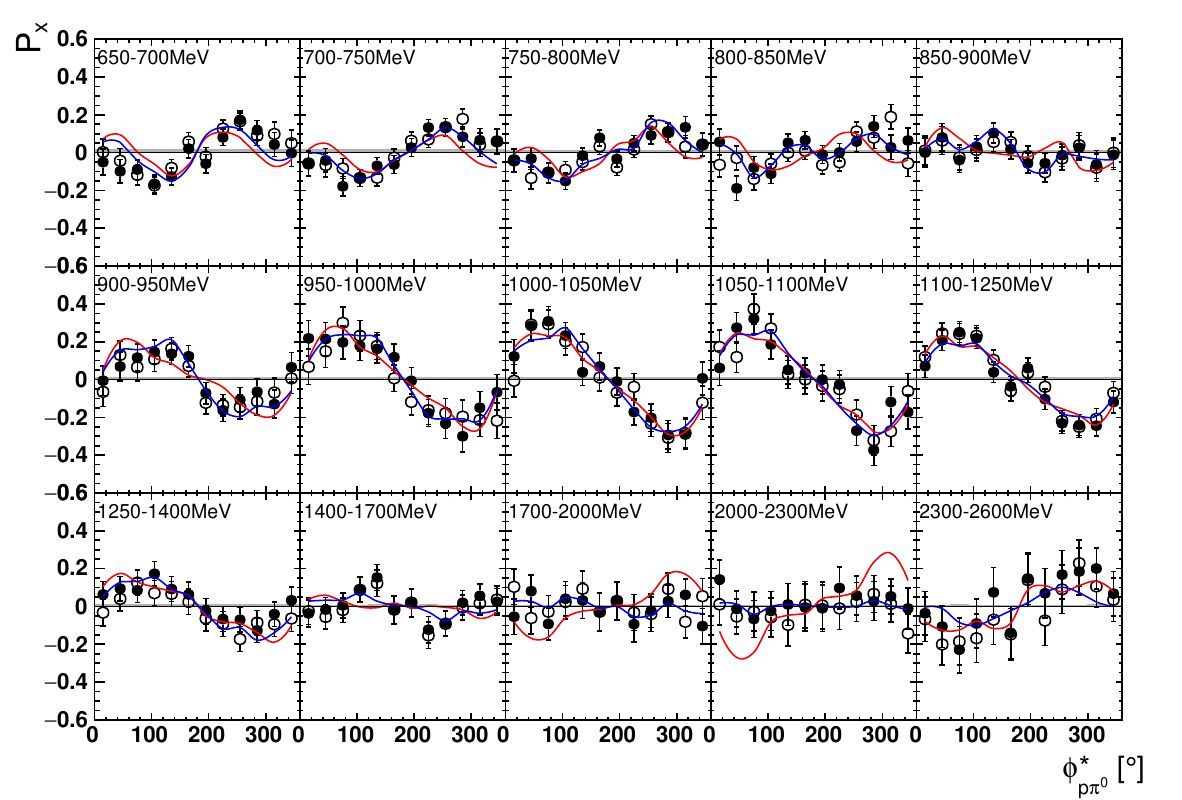}
} \caption{The target asymmetry $\rm P_x$ as a function of beam energy $E_{\Pgg}$ and $\phi^*_{\Pp\Pgpz}$. The
open symbols make use of the symmetry properties. The colored lines represent PWA solutions: BnGa 2014-02 in red, new BnGa 2022-02 in blue. The systematic uncertainty is shown as a gray band.}
\label{fig:Px_ppi0}
\end{center}
\end{figure*}

\begin{figure*}
\begin{center}
\resizebox{.85\textwidth}{!}{%
  \includegraphics{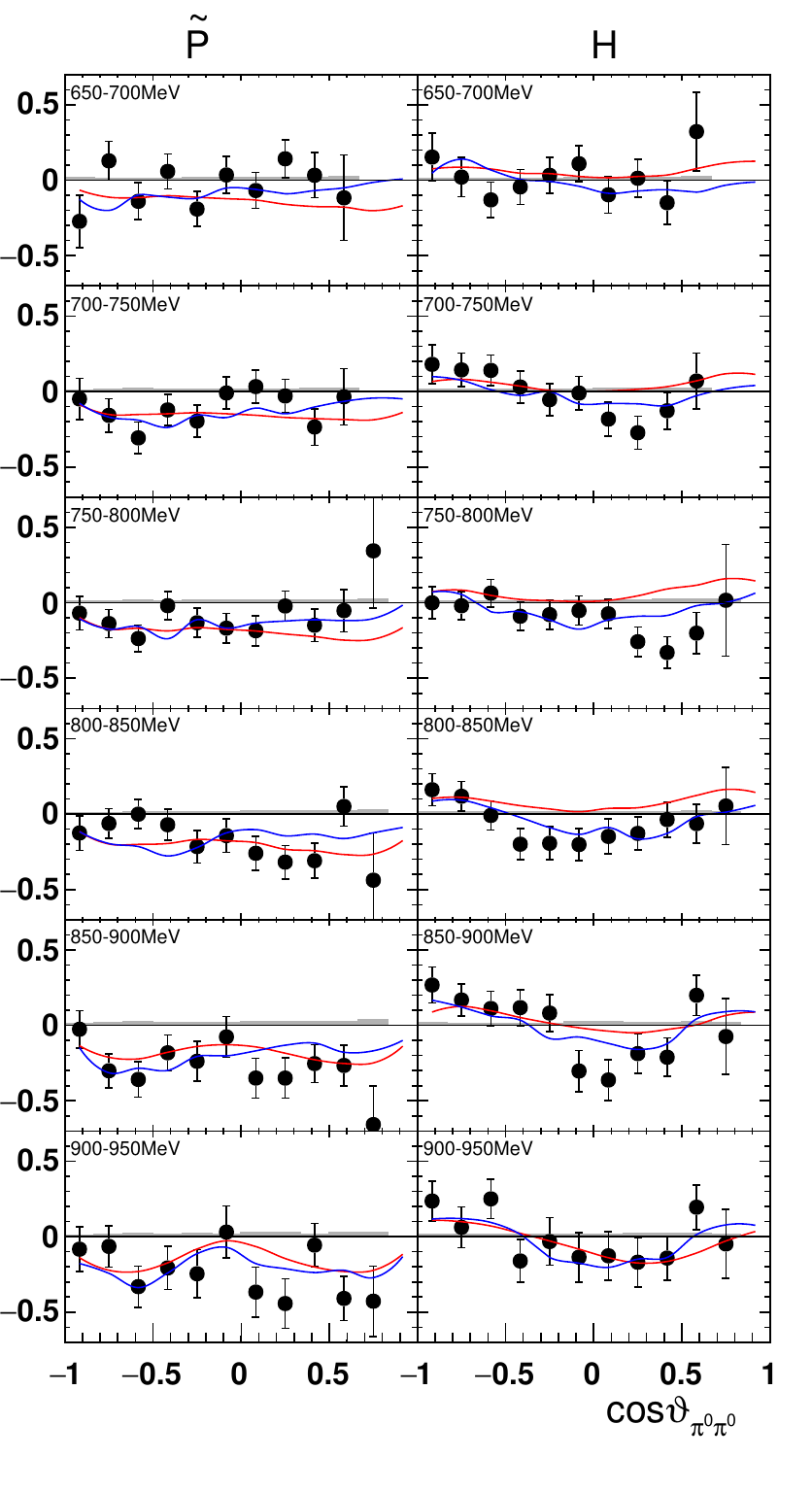}
  \includegraphics{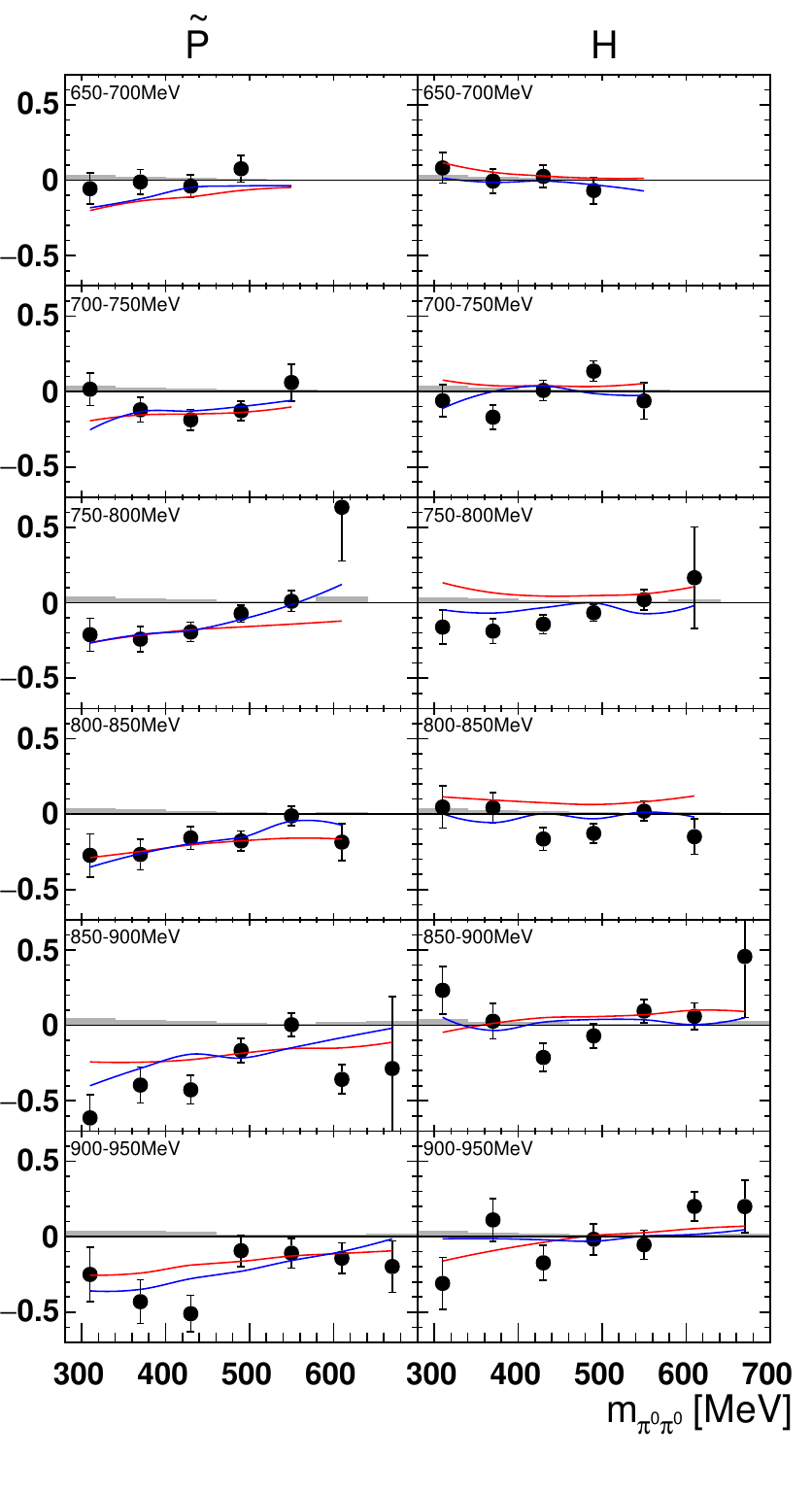}
} \caption{The double polarization observables $\rm\tilde P$ and $\rm H$ as a function of beam energy $E_{\Pgg}$ and $\cos\vartheta_{\Pgpz\Pgpz}$ (left), or $E_{\Pgg}$ and $m_{\Pgpz\Pgpz}$ (right). The colored lines represent PWA solutions: BnGa 2014-02 in red, new BnGa 2022-02 in blue. The systematic uncertainty is shown as a gray band.}
\label{fig:PH}
\end{center}
\end{figure*}

\begin{figure*}
\begin{center}
\resizebox{.85\textwidth}{!}{%
  \includegraphics{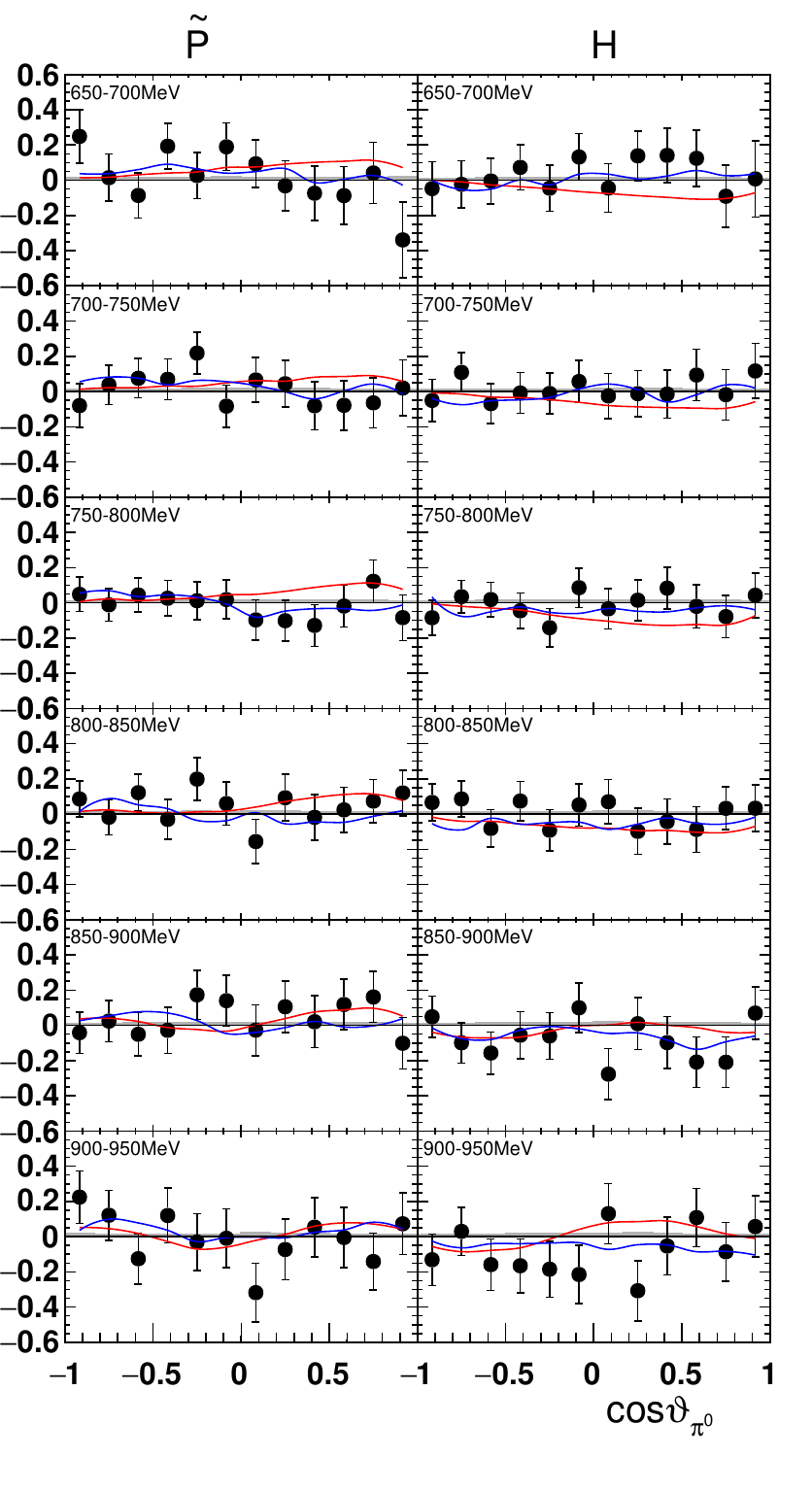}
  \includegraphics{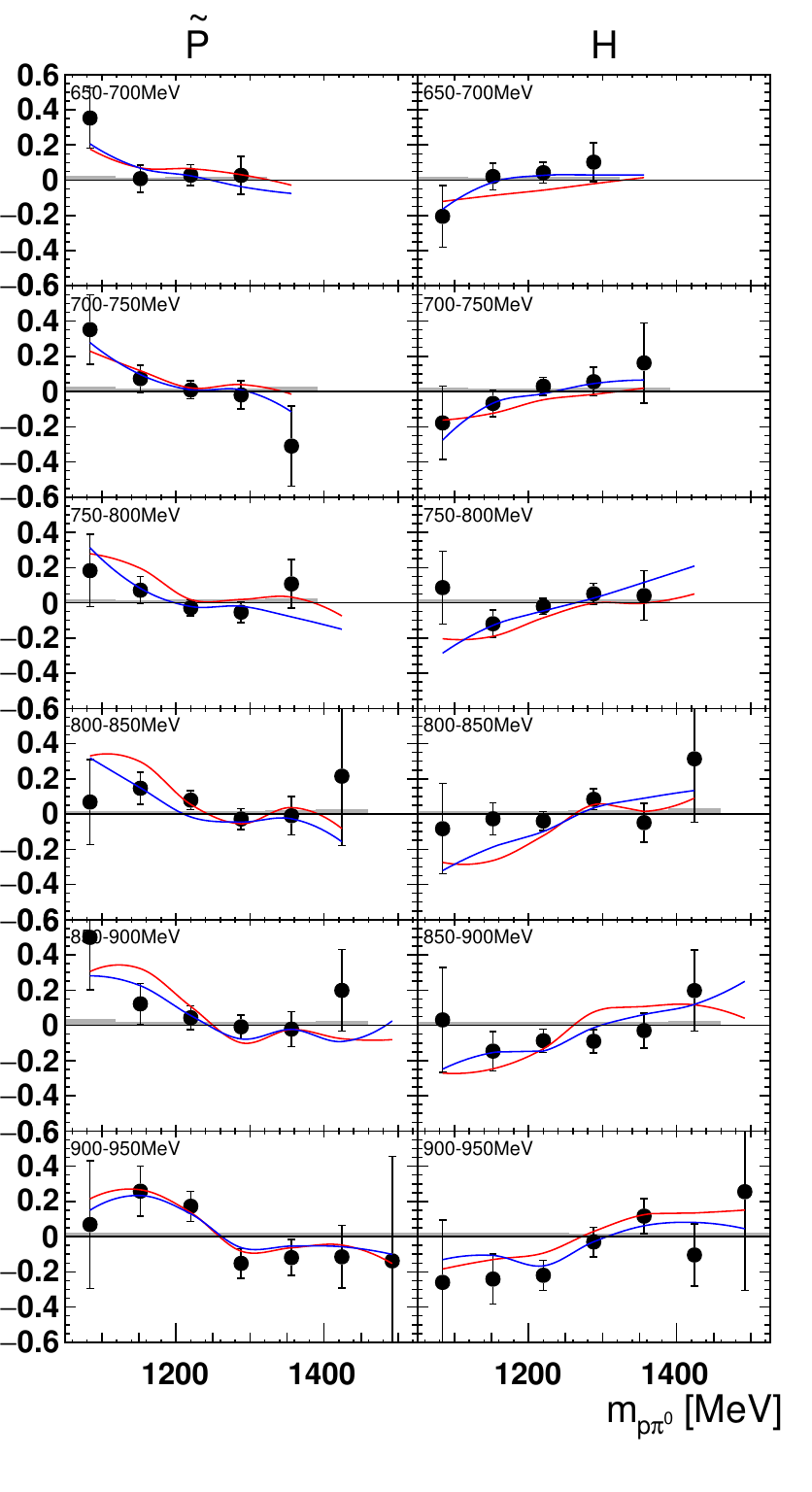}
} \caption{The double polarization observables $\rm\tilde P$ and $\rm H$ as a function of beam energy $E_{\Pgg}$ and $\cos\vartheta_{\Pgpz}$ (left), or $E_{\Pgg}$ and $m_{\Pp\Pgpz}$ (right). The colored lines represent PWA solutions: BnGa 2014-02 in red, new BnGa 2022-02 in blue. The systematic uncertainty is shown as a gray band.}
\label{fig:PH_ppi0}
\end{center}
\end{figure*}

\subsection{Results}
\label{sec:results}
Due to the limited statistics, a full analysis in 5-dimensional bins could not be performed. Instead, a
series of 2-dimen\-sional analyses were done where all pair-wise combinations, \textit{i.e.}

\hspace{-5mm}\begin{tabular}{lll}
$E_{\Pgg}$ -- $\cos\vartheta_{\Pgpz\Pgpz}$, & $E_{\Pgg}$ -- $m_{\Pgpz\Pgpz}$,& $\cos\vartheta_{\Pgpz\Pgpz}$ -- $m_{\Pgpz\Pgpz}$,\\
$E_{\Pgg}$ -- $\theta^*_{\Pgpz\Pgpz}$,& $E_{\Pgg}$ -- $\phi^*_{\Pgpz\Pgpz}$,& $\cos\vartheta_{\Pgpz\Pgpz}$ -- $\phi^*_{\Pgpz\Pgpz}$, \dots
\end{tabular}
were investigated.

A selection of results is shown in Figs.~\ref{fig:T_costh}-\ref{fig:Px_ppi0} for the target asymmetries and in Figs.~\ref{fig:PH} and \ref{fig:PH_ppi0} for the double polarization observables together with predictions from the $2\Pgp$-MAID~\cite{maid}, BnGa2014-02~\cite{bnga} and the new fit of the BnGa PWA.

In case the observable is plotted versus $\cos\theta_{\Pgpz\Pgpz}$, $m_{\Pgpz\Pgpz}$, or $\phi^*_{\Pgpz\Pgpz}$, this means that the two pions span the decay plane (particles 2 and 3 in Fig.~\ref{fig:3body}). On the other hand, if $\cos\theta_{\Pgpz}$, $m_{\Pp\Pgpz}$, or $\phi^*_{\Pp\Pgpz}$ are used, the proton and one of the pions are particles 2 and 3.

While some deviations occur, the BnGa2014-02 describes the data reasonably well at lower energies. But at higher energies, especially above \unit[1700]{MeV}, larger discrepancies become visible. The $2\Pgp$-MAID prediction is rarely consistent with the data.

In case of the observables being plotted against $\phi^*$, several symmetry properties arise.
Observables that are even under the transformation $\phi^*\mapsto2\pi-\phi^*$ do not vanish while integrating over $\phi^*$, while the others do.
Additionally, the two pions are indistinguishable, which is why the observables are invariant under 
exchange of the pions. Thus, if the two pions are spanning the decay plane (vectors $p_2$ and $p_3$ in
Fig.~\ref{fig:3body}) the observables are unchanged by the transformation $\phi^*\mapsto\phi^*+\pi$.
This symmetry property is fulfilled exactly in the data, since the pions were symmetrized.
Data points calculated using these symmetry properties are represented as open symbols in Figs.~\ref{fig:Py}, \ref{fig:Px},
\ref{fig:Py_ppi0} and \ref{fig:Px_ppi0}. The good agreement between the closed and open symbols
(their difference\footnote{When plotting against $\phi^*_{\Pgpz\Pgpz}$ there are only 3 independent bins (with 12 $\phi^*$-bins) per energy bin due to the symmetrization (indistinguishable pions).} normalized to their statistical uncertainty follows a standard normal distribution with mean $\mu=0.04\pm0.06$ and $\sigma=1.01\pm0.04$)
shows that the symmetry properties are indeed fulfilled in the data.

\begin{figure*}
\begin{center}
\resizebox{.823\textwidth}{!}{%
  \includegraphics{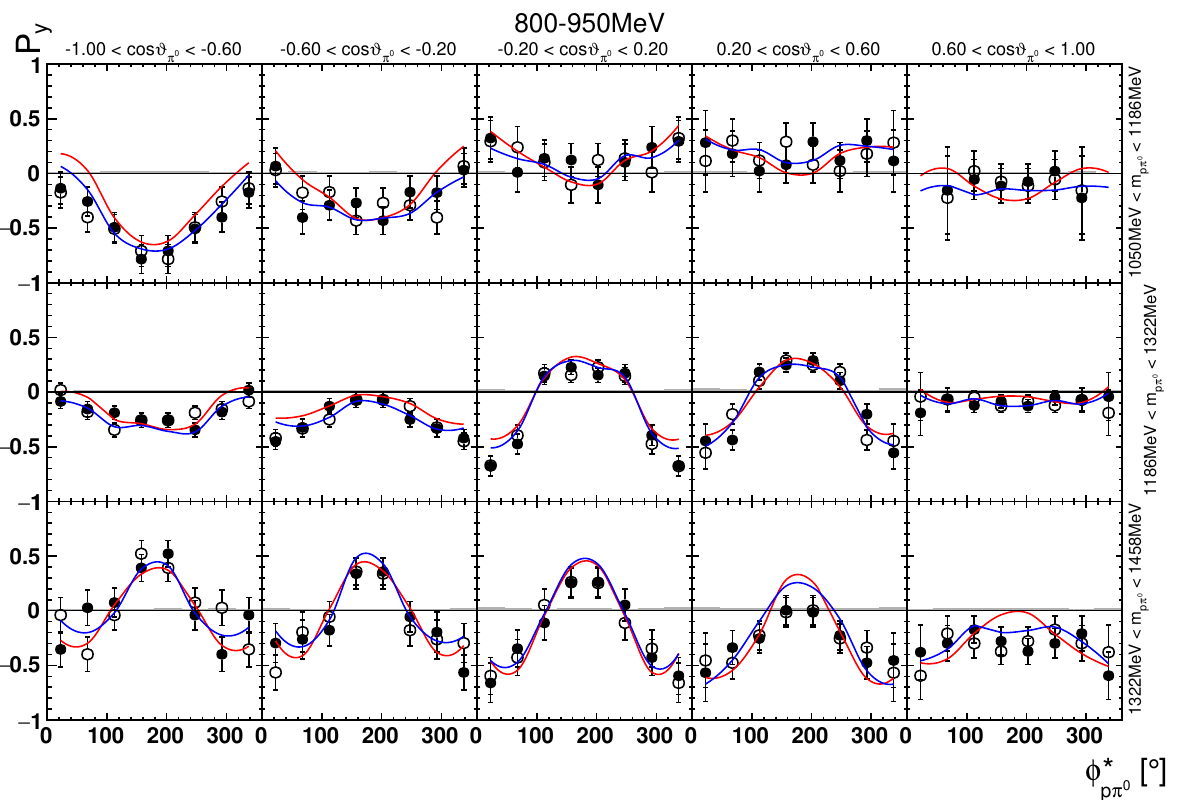}
} \caption{Four-dimensional determination of the target asymmetry $\rm P_y$ as a function of $\phi^*_{\Pp\Pgpz}$.
The $\cos\vartheta_{\Pgpz}$ is varied within a row, the invariant mass $m_{\Pp\Pgpz}$ within a column.
Only a single energy bin is shown here. The colored lines represent PWA solutions: BnGa 2014-02 in red, new BnGa 2022-02 in blue. The systematic uncertainty is shown as a gray band.}
\label{fig:Py_4D}
\end{center}
\begin{center}
\resizebox{.823\textwidth}{!}{%
  \includegraphics{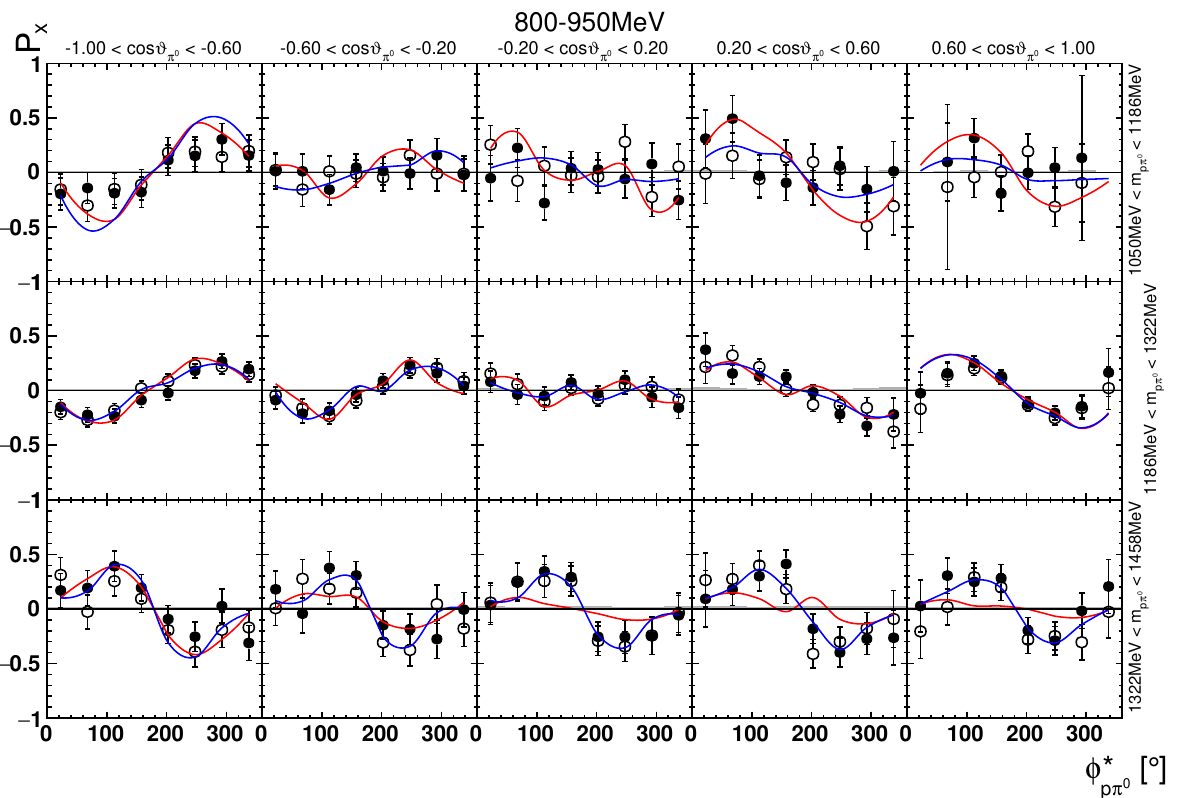}
} \caption{Four-dimensional determination of the target asymmetry $\rm P_x$ as a function of $\phi^*_{\Pp\Pgpz}$.
 The $\cos\vartheta_{\Pgpz}$ is varied within a row, the invariant mass $m_{\Pp\Pgpz}$ within a column.
 Only a single energy bin is shown here. The colored lines represent PWA solutions: BnGa 2014-02 in red, new BnGa 2022-02 in blue. The systematic uncertainty is shown as a gray band.}
\label{fig:Px_4D}
\end{center}
\end{figure*}

\begin{figure*}
\begin{center}
\resizebox{.823\textwidth}{!}{%
  \includegraphics{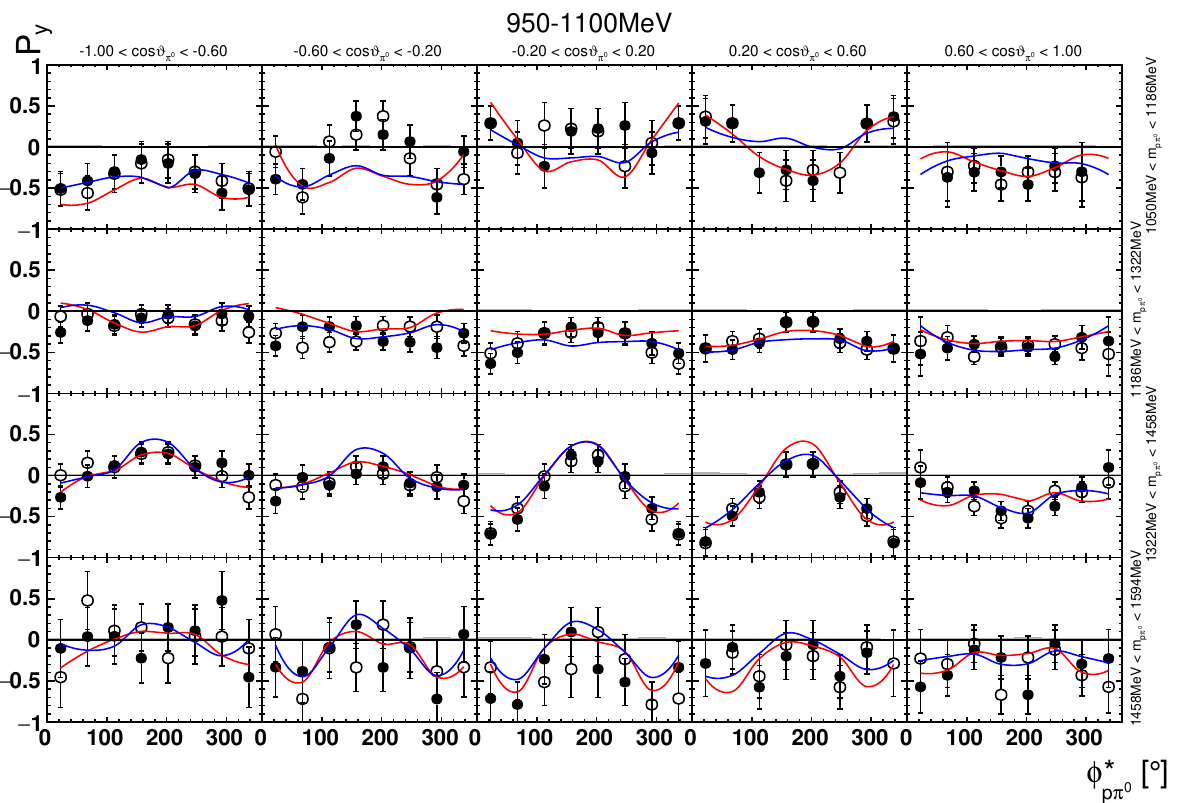}
} \caption{Four-dimensional determination of the target asymmetry $\rm P_y$ as a function of $\phi^*_{\Pp\Pgpz}$.
 The $\cos\vartheta_{\Pgpz}$ is varied within a row, the invariant mass $m_{\Pp\Pgpz}$ within a column.
 Only a single energy bin is shown here. The colored lines represent PWA solutions: BnGa 2014-02 in red, new BnGa 2022-02 in blue. The systematic uncertainty is shown as a gray band.}
\label{fig:Py_4D2}
\end{center}
\begin{center}
\resizebox{.823\textwidth}{!}{%
  \includegraphics{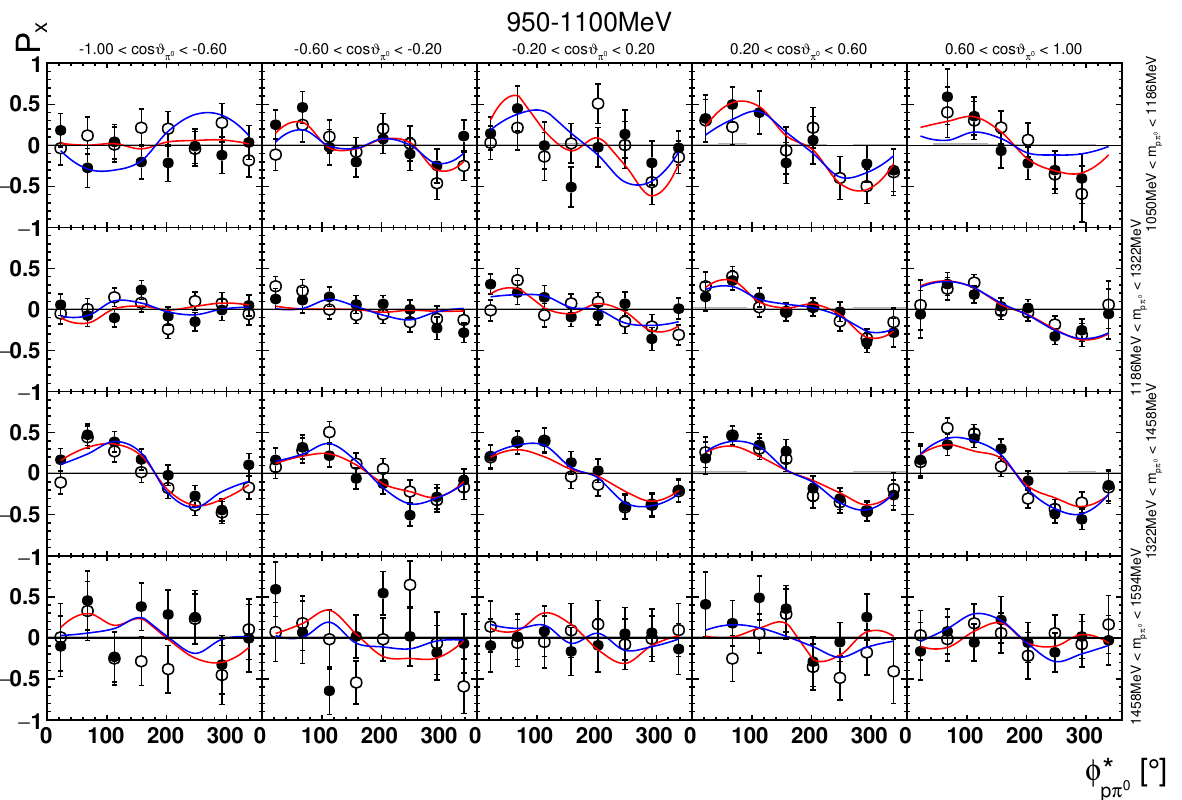}
} \caption{Four-dimensional determination of the target asymmetry $\rm P_x$ as a function of $\phi^*_{\Pp\Pgpz}$.
 The $\cos\vartheta_{\Pgpz}$ is varied within a row, the invariant mass $m_{\Pp\Pgpz}$ within a column.
 Only a single energy bin is shown here. The colored lines represent PWA solutions: BnGa 2014-02 in red, new BnGa 2022-02 in blue. The systematic uncertainty is shown as a gray band.}
\label{fig:Px_4D2}
\end{center}
\end{figure*}

In addition to the 2-dimensional analyses, a 4-dimen\-sional analysis was performed for the
target asymmetries below \unit[1250]{MeV}. Here, the beam energy $E_{\Pgg}$, $\cos\vartheta_{\Pgpz}$,
$m_{\Pp\Pgpz}$ as well as $\phi^*_{\Pp\Pgpz}$ were used as kinematic variables, only $\theta^*_{\Pp\Pgpz}$ was
integrated out. Fig.~\ref{fig:Py_4D} shows an example of this analysis where one energy bin of
the target asymmetry $\rm P_y$ is plotted as a function of $\phi^*_{\Pp\Pgpz}$. The $\cos\vartheta_{\Pgpz}$ is
varied in the rows while $m_{\Pp\Pgpz}$ is varied in the columns. Fig.~\ref{fig:Py_4D2} shows the same for a higher energy bin. The analogue result for the
target asymmetry $\rm P_x$ can be found in Fig.~\ref{fig:Px_4D} and Fig.~\ref{fig:Px_4D2}, respectively. In all cases the aforementioned
symmetry properties are fulfilled very well.

It is easily seen that the observables can be significantly different in different bins. For
example, the lowest $\cos\vartheta_{\Pgpz}$- and $m_{\Pp\Pgpz}$-bin in Fig.~\ref{fig:Py_4D} shows the
opposite behavior in $\phi^*_{\Pp\Pgpz}$ from the middle $\cos\vartheta_{\Pgpz}$- and $m_{\Pp\Pgpz}$-bin.

\begin{figure*}
\resizebox{\textwidth}{!}{%
  \includegraphics{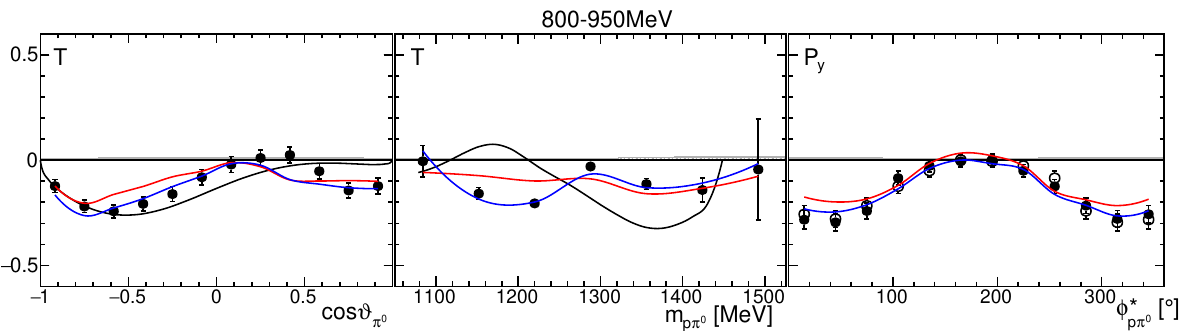}
} \caption{Partially integrated target asymmetry for the same energy bin as shown in
Fig.~\ref{fig:Py_4D} as a function of $\cos\vartheta_{\Pgpz}$ (left), $m_{\Pp\Pgpz}$ (center), and
$\phi^*_{\Pp\Pgpz}$ (right). The colored lines represent PWA solutions: $2\pi$-MAID in black, BnGa 2014-02 in red, new BnGa 2022-02 in blue. The systematic uncertainty is shown as a gray band.}
\label{fig:Py_4D_integ}
\end{figure*}

By comparing the 4-dimensional result to the partially integrated ones (shown in
Fig.~\ref{fig:Py_4D_integ} for the same energy bin as depicted in Fig.~\ref{fig:Py_4D}) it becomes
immediately clear that information is lost in the integrated observables.

\section{\label{sec:PWA}Partial wave analysis}
The data on the target asymmetries and double polarization observables for the reaction $\Pgg\Pp\to\Pp\Pgpz\Pgpz$ were included in the
data base for the Bonn-Gatchina (BnGa) multi-channel partial-wave analysis (PWA), one of the current state of the art PWAs (see \cite{Thiel:2022xtb,Doring:2025sgb} and references therein).

The formalism used in the BnGa-PWA is described in detail in
Refs.~\cite{Anisovich:2004zz,Anisovich:2006bc,Anisovich:2007zz,Denisenko:2016ugz}. In addition to elastic and inelastic $\Pgp\PN$-scattering data, the analysis
includes data on photo-induced reactions with two and three particles in the final state like
$\Pgg \Pp\to \Pp\Pgpz$, $\Pn\Pgpp$, $\Pp\Pgh$, $\Pp\Pghpr$, $\Pp\Pgo$, $\Pp\Pgpz\Pgpz$, $\Pp\Pgpz\Pgh$, $\PgL \PKp$, $\PgSz \PKp$, $\PgSp \PKz$, and $\Pp\Pgpp\Pgpm$.
The inclusion of data on the latter reaction is important. We illustrate this importance using $\PN(1700)3/2^-$
and $\PgD(1700)3/2^-$ as examples. The reaction $\Pgg \Pp\to \PKp\PgL$ or $\Pgg \Pp\to \Pgh \Pp$
can proceed only via $\PN(1700)3/2^-$ formation, while $\Pgpp\Pp\,\to\,\Pgpp\Pp$\, or\, $\PKp\PgSp$
proceeds only via $\PgD(1700)3/2^-$. The two resonances can be separated in $\Pgpm\Pp$ elastic and
charge exchange scattering due to their different Clebsch-Gordan (CG) coefficients to $\Pp\Pgpz$
and $\Pn\Pgpp$. There is, however, no information to separate contributions from $\PN(1700)3/2^-$ and
$\PgD(1700)3/2^-$ in the reaction $\Pgg \Pp\to \Pp\Pgpz\Pgpz$. The same is of course true for the higher mass $\PN^*$- and $\PgD^*$-resonances. Technically, the amplitudes for the two isospins
have the tendency to become very large in the PWA and to interfere destructively. In
Ref.~\cite{Sokhoyan:2015fra} this was avoided by penalizing, \textit{i.e.} limiting the photo-couplings of large amplitudes interfering
destructively. With data on $\Pgg\Pp\to\Pp\Pgpp\Pgpm$ in the fit, the contributions of $\PN^*$'s and
$\PgD^*$'s can be separated due to their different CG couplings.

Pion-induced reactions are included as well, in particular the real and imaginary part of the $\Pgp
\PN$ elastic scattering partial-wave amplitudes from the GWU analysis~\cite{Arndt:2006bf} or,
alternatively, from the Karlsruhe-Helsinki analysis (KH) \cite{Hohler:1979yr}. New data have been
included in the BnGa data base, \textit{e.g.} (double) polarization data on $\Pgg\Pp\to\Pp\Pgh$ \cite{mueller} and data on the reaction $\Pgg\Pp\to\Pp\Pgpp\Pgpm$. Data on the latter reaction were
published recently \cite{Golovatch:2018hjk}, where reaction cross section and nine one-dimensional
histograms were reported. Data on this reaction were made available to us on a single-event basis.
Thus, the data could be used in an event-by-event likelihood fit.

The BnGa PWA uses background terms -- represented by $t$- and $u$-channel exchange processes
and by a few mostly constant terms added to the $K$-matrix pole terms -- and a large number of resonances
to describe the resonant part of meson-baryon interactions. Resonances are coupled to all allowed
decay channels. When branching ratios turned out to be very small ($\lesssim 1$\%) the couplings
were set to zero to avoid an excessive number of parameters.

For all data with two-body final states or polarization data in 3-body final states, the $\chi^2$ of the fit is calculated. Unpolarized data with three
particles in the final state are fitted event-by-event in a likelihood fit. This method guarantees
optimum use of all correlations between the different variables the event depends on. The fit minimizes the pseudo-likelihood
defined by
\begin{equation}
 -\ln {\cal L}_{\rm tot}= \left( \frac 12\sum_i w_i\chi^2_i-\sum_i w_i\ln{\cal L}_i \right) \ \frac{\sum_i
N_i}{\sum_i w_i N_i} \label{likeli} .
\end{equation}
Data sets are given a weight. Without weights, low-statistics data, \textit{e.g.} on polarization observables,
may be reproduced unsatisfactorily without significant deterioration of the total ${\cal L}_{\rm
tot}$. For any data set, its weight is increased as long as the gain in likelihood for the data set
is larger than twice the loss in the overall likelihood. The total likelihood function is
normalized to avoid an artificial increase in statistics by the weighting factors. Differences in
the fit quality are given as $\chi^2$ difference, with $\Delta\chi^2=-2\Delta{\ln\cal L}_{\rm tot}$.

The main impact of the data presented here is in the observation of cascade decays into
$\Pp\Pgp\Pgp$ via intermediate resonances. Table~\ref{tab:Decayall} shows the results. In the
fit, the intermediate resonances listed in the table were admitted. The large number of
contributing resonances and decay modes leads to a large number of close-by solutions resulting in
the uncertainties given in the table. The new numbers are compared to those from \cite{Sokhoyan:2015fra}. Most
of the branching ratios are compatible with the previous findings within $1\sigma$;
the difference between the new and the older results exceeds $2\sigma$ only for a few exceptions.

\begin{table*}
\caption{\label{tab:Decayall}Branching ratios (in \%) for decays of nucleon and $\PgD^*$
resonances. The spread of results from different solutions is used to estimate the uncertainties.
x signifies forbidden transitions. Entries marked as ``-''  were fitted to small values and set
to zero to avoid an excessive number of free parameters. The intermediate resonances on the right
side of the double line carry orbital excitations. Where available, the results from Ref.~\cite{Sokhoyan:2015fra}
are given as small numbers. For $\PN(1710)1/2^+$ two values were given in~\cite{Sokhoyan:2015fra};
both are reproduced here. Small numbers marked with $\dagger$ were taken from \cite{Hunt}. An inclusion of the
$\PgD(1750)1/2^+$ improves the fit quality only slightly, its existence cannot be established from our fits.
Allowing for $\PgD(2000)3/2^+\to\PN(1700)3/2^-\Pgp$ decays leads to an improvement of the fit. Due to its large
uncertainty, the respective branching ratio is not given in the table.}
\begin{center}
\renewcommand{\arraystretch}{1.38}
{\scriptsize
\begin{tabular}{|c|l|c|cc|cc||c|c|c|c|c|} \hline
                 && \ $\PN\pi$\quad$L$&$\PgD\pi$&\LlJ&$\PgD\pi$ &\LgJ&$\PN(1440)\pi$ $L$   &$\PN(1520)\pi$ $L$ &$\PN(1535)\pi$ $L$    &$\PN(1680)\pi$ $L$ &$\PN\sigma$\quad $L$\\ \hline
\parbox[t]{3mm}{\multirow{12}{*}{\rotatebox[origin=c]{90}{1st shell, one oscillator}}}
     &$\PN(1535)1/2^-$&46\er5     \hf0&   x       &    &5\er3   & 2  &\ \ 6\er5      \hf0&\ \ \quad -    \hf 1&\ \ \quad -        \hf1&\ \ \quad -      \hf2&    4\er2     \hf1\\[-1ex]
                   &&\ty 52\er 5\hp0&           &    &\ty2.5\er1.5&    &\ \ty\ 12\er8  \hp1&\ \ \quad -    \hp 1&\ \ \quad -        \hp2&\ \ \quad -      \hp2&\ty 6\er 4    \hp0\\[-0.5ex]
     &$\PN(1520)3/2^-$&61\er3     \hf2&   10\er 4 & 0  &     10\er 3 & 2  &\ \ $<$1       \hf2&\ \ \quad -    \hf 1&\ \ \quad -      \hf1&\ \ \quad -      \hf2&$<$2         \hf1\\[-1ex]
                   &&\ty 61\er 2\hp0&\ty19\er 4 &    &\ty  9\er 2 &    &\ \ \ty $<$1   \hp1&\ \ \quad -    \hp 1&\ \ \quad -        \hp2&\ \ \quad -      \hp2&\ty$<2$       \hp0\\[-0.5ex]
     &$\PN(1650)1/2^-$&48\er4     \hf0&   x       &    &    6\er 3  & 2  &\ \ 5\er3    \hf0&\ \ \quad -    \hf 1&\ \ \quad -        \hf1&\ \ \quad -      \hf2& 3\er2        \hf1\\[-1ex]
                   &&\ty 51\er 4\hp0&           &    &\ty 12\er 6 &    &\ \ \ty16\er10  \hp1&\ \ \quad -    \hp 1&\ \ \quad -        \hp2&\ \ \quad -      \hp2&\ty  10\er 8  \hp0\\[-0.5ex]
     &$\PN(1700)3/2^-$&20\er 8     \hf2&   66\er 17 &  0 &   7\er4    & 2  &\ \ 9\er5      \hf2& \quad$<$2  \hf 1&\quad$<$1       \hf1&\ \ \quad -     \hf2&6\er4        \hf1\\[-1ex]
                   &&\ty 15\er 6\hp0&\ty65\er 15&    &\ty  9\er 5 &    &\ \ \ty  7\er4 \hp2&\quad\ty$<$4\hp 1&\quad\ty$<$1    \hp1&\ \ \quad -     \hp2&\ty  8\er 6   \hp0\\[-0.5ex]
     &$\PN(1675)5/2^-$&40\er1     \hf2&   19\er3   & 2  &\quad -     & 4  &\ \ \quad -\     \hf2&\ \ \quad -    \hf 1&\ \ \quad -        \hf3&\ \ \quad -     \hf0& 1\er1        \hf3\\[-1ex]
                   &&\ty 41\er 2\hp0&\ty30\er 7 &    &\quad -     &    &\ \ \quad -      \hp2&\ \ \quad -    \hp 1&\ \ \quad -        \hp3&\ \ \quad -     \hp0&\ty  5\er 2   \hp0\\[-0.5ex]
 &$\PgD(1620)1/2^-$&30\er5    \hf0&    x      &    &    28\er15 & 2  &\ \ 15\er8     \hf0&\ \ \quad -    \hf 1&\ \ \quad -        \hf1&\ \ \quad -     \hf2&\quad x      \hp0\\[-1ex]
                   &&\ty 28\er 3\hp0&           &    &\ty 62\er10 &    &\ \ \ty 6\er  3\hp0&\ \ \quad -    \hp 1&\ \ \quad -        \hp1&\ \ \quad -     \hp2&                  \\[-0.5ex]
 &$\PgD(1700)3/2^-$&22\er6    \hf2&   16\er15 & 0  &   8\er6    & 2  & \ \ 3\er2     \hf2&\quad $<$1 \hf 1&\quad $<$1    \hf1&\ \ \quad -     \hf2&\quad x      \hp0\\[-1ex]
                   &&\ty 22\er 4\hp0&\ty 20\er15&    &\ty 10\er 6 &    &\ \ \ty $<$1   \hp2&\quad\ty3\er2\hp1&\quad\ty$<$1    \hp1&\ \ \quad -     \hp2& \\[-0.5ex]
\hline\hline
\parbox[t]{3mm}{\multirow{10}{*}{\rotatebox[origin=c]{90}{2nd shell, one oscillator}}}
     &$\PN(1440)1/2^+$&66\er 3    \hf1&    x      &    &     10\er6& 1  &\ \  \quad  -\hf1&\ \ \quad  -      \hf0&\ \  \quad -       \hf2&\ \ \quad -     \hf3&  17\er6      \hf1\\[-1ex]
                   &&\ty 63\er2 \hp0&           &    &\ty 20\er7&    &\ \ \quad -\ty\   \hp1&\ \ \quad -    \hp 1&\ \  \quad -     \hp2&\ \ \quad -      \hp2&\ty 17\er6    \hp0\\[-0.5ex]
     &$\PgD(1600)3/2^+$& 17\er4     \hf1&   70\er6   & 1  &     $<$2& 3  & \ \ $<$1     \hf1&\ \ \quad -   \hf0&\ \  \quad -      \hf2&\ \ \quad -     \hf1&\quad x      \hp0\\[-1ex]
                   &&\ty 14\er  4\hp1&\ty 77\er5&    &\ty  $<$2&    & \ \ \ty 22\er5$^\dagger$   \hp1&\ \ \quad - \hp0&\ \ \quad -    \hp2&\ \ \quad -     \hp1&                  \\[-0.5ex]
     &$\PN(1720)3/2^+$&13\er5     \hf1&    15\er7& 1  &      6\er6 & 3  & \ \ 6\er5      \hf1& \ \ 7\er3  \hf0& \ \ 4\er2       \hf2&\ \ \quad -     \hf1&    20\er10     \hf2\\[-1ex]
                   &&\ty11\er  4\hp1&\ty 62\er15&    &\ty   6\er 6&    & \ \ \ty$<$2   \hp1& \ \ \ty3\er2 \hp0& \ \ \ty$<$2    \hp2&\ \ \quad -     \hp1&\ty 8\er6     \hp1\\[-0.5ex]
     &$\PN(1680)5/2^+$&68\er 8    \hf3&    8\er4 & 1  &     8\er4  & 3  &\ \ \quad-         \hf3& \ \ $<$1    \hf2&\ \ \quad -         \hf2&\ \ \quad -     \hf1&   8\er 4    \hf2\\[-1ex]
                   &&\ty62\er  4\hp3&\ty 7\er 3 &    &\ty  10\er 3&    &\ \ \quad-         \hp3& \ \ \ty $<$1\hp2&\ \ \quad -         \hp2&\ \ \quad -     \hp1&\ty14\er 5    \hp3\\[-0.5ex]
&$\PgD(1910)1/2^+$&16\er 6    \hf1&    x      &    &     17\er9& 1  & \ \    40\er15\hf1&\ \ \quad  -      \hf0& \ \ 4\er2      \hf2&\ \ \quad -     \hf3& \quad x      \hp0\\[-1ex]
                   &&\ty12\er 3\hp1&           &    &\ty  50\er16&    & \ \ \ty6\er 3 \hp1&\ \ \quad -      \hp0& \ \ \ty5\er3   \hp2&\ \ \quad-      \hp3&                   \\[-0.5ex]
&$\PgD(1920)3/2^+$& 12\er6     \hf1&   5\er4   & 1  &     40\er20& 3  & \ \ 9\er6     \hf1& \ \ 10\er8   \hf0& \ \ 5\er5      \hf2&\ \ \quad -     \hf1&\quad x      \hp0\\[-1ex]
				   &&\ty 8\er  4\hp1&\ty18\er 10&    &\ty  58\er14&    & \ \ \ty$<4$   \hp1& \ \ \ty$<5$ \hp0& \ \ \ty$<2$    \hp2&\ \ \quad -     \hp1&                  \\[-0.5ex]
&$\PgD(1905)5/2^+$&13\er 4    \hf3&  20\er12   & 1  &     -      & 3  &\ \ \quad -        \hf3&\ \ \quad -      \hf2& \ \ $<1$       \hf2& \ \ 6\er2  \hf1&\quad x      \hp0\\[-1ex]
                   &&\ty13\er  2\hp3&\ty33\er 10&    &     -      &    &\ \ \quad -        \hp3&\ \ \quad -      \hp2& \ \ \ty$<1$    \hp2& \ \ \ty10\er5\hp1&                  \\[-0.5ex]
&$\PgD(1950)7/2^+$&46\er 4    \hf3&    5\er4  &3   &     -      & 5  &\ \ \quad -        \hf3&\ \ \quad -      \hf2&\ \ \quad -         \hf4& \ \ 3\er2   \hf1&\quad x      \hp0\\[-1ex]
                   &&\ty46\er  2\hp3&\ty 5\er  4&    &     -      &    &\ \ \quad -        \hp3&\ \ \quad -      \hp2&\ \ \quad -         \hp4& \ \ \ty 6\er3\hp1&                  \\[-0.5ex]
\hline\hline
\parbox[t]{3mm}{\multirow{10}{*}{\rotatebox[origin=c]{90}{2nd shell, mixed oscillations}}}
& $\PN(1710)1/2^+$    & 5\er 3   \hf1&    x       &    &     6\er4& 1  & \ \   22\er12    \hf1& \ \ $<$2     \hf2& \ \ 4\er4   \hf0&\ \ \quad -     \hf3& 14\er 6\hf0\\[-1ex]
                   && \ty5\er 3\hp0&            &    &\ty   7\er 4&    & \ \ \ty30\er10\hp1& \ \ \ty$<$2   \hp2&\ \ \quad -       \hp0&\ \ \quad -     \hp3& \ty55\er15\hp0\\[-1ex]
                   && \ty5\er 3\hp0&            &    &\ty  25\er10&    & \ \ \ty$<$5   \hp1& \ \ \ty$<$2   \hp2& \ \ \ty15\er6\hp0&\ \ \quad -     \hp3& \ty10\er5\hp0\\[-0.5ex]
&\it $\PgD\it(1750)1/2^+$&18\er 5   \hf1&    x       &    &     22\er10& 1  & \ \   49\er 26\hf1&\ \ \quad  -        \hf0& \ \ $<$2  \hf2&\ \ \quad -     \hf3& \quad x      \hp0\\[-1ex]
                   && \hp1         &           &    &            &    & \ \ \ty    \hp1&\ \ \quad         \hp0& \ \ty  \hp2&\quad      \hp3&                  \\[-0.5ex]
&$\PgD(2000)3/2^+$& 11\er5     \hf1&   14\er6   & 1  &     3\er2& 3  & \ \ 9\er6     \hf1& \ \ $<$1   \hf0& \ \ $<$1      \hf2&\ \ \quad -     \hf1&\quad x      \hp0\\[-1ex]
                   && \hp1&&    &&    & \ \    \hp1& \ \  \hp0& \ \     \hp2&      \hp1&                  \\[-0.5ex]
 &$\PN(2100)1/2^+$    &17\er 7   \hf1&    x       &    &     12\er6& 1  &\ \ \quad -        \hf1& \ \ $<$2      \hf2& \ \ $<$1   \hf0&\ \ \quad -     \hf3&28\er7\hf0\\[-1ex]
                   &&\ty16\er 5\hp0&            &    &\ty  10\er 4&    &\ \ \quad -        \hp1& \ \ \ty$<$2   \hp2& \ \ \ty30\er4\hp0&\ \ \quad -     \hp3& \ty20\er 6\hp0\\[-0.5ex]
 &$\PN(1880)1/2^+$    &   11\er6  \hf1&    x      &    &     4\er2& 1  &\ \ \quad -        \hf1&\ \ \quad -      \hf2& \ \ 6\er3      \hf0&\ \ \quad -     \hf3&22\er13 \hf0\\[-1ex]
                   && \ty6\er  3\hp0&           &    &\ty  30\er12&    &\ \ \quad -        \hp1&\ \ \quad -      \hp2& \ \ \ty8\er4   \hp0&\ \ \quad -     \hp3&\ty25\er 15\hp0\\[-0.5ex]
 &$\PN(1900)3/2^+$    &    4\er3  \hf1&   9\er6  & 1  &     4\er3& 3  & \ \ 9\er6     \hf1& \ \ $<$2 \hf0& \ \ 15\er 6     \hf2&\ \ \quad -     \hf1&9\er 4\hf2\\[-1ex]
                   && \ty3\er  2\hp0&\ty17\er 8 &    &\ty  33\er12&    & \ \ \ty$<$2   \hp1& \ \ \ty15\er8\hp0& \ \ \ty7\er3   \hp2&\ \ \quad -    \hp1&\ty4\er 3\hp0\\[-0.5ex]
     &$\PN(1860)5/2^+$&12\er4    \hf3&    2\er2 & 1  &     4\er2  & 3  &\ \ \quad-         \hf3&\ \  \quad-    \hf2&\ \ \quad -         \hf2&\ \ \quad -     \hf1&   4\er4    \hf2\\[-1ex]
                   &&\ty20\er  6\hp3&\ty  10\er6$^\dagger$ &    &\ty  27\er11$^\dagger$ &    &\ \ \quad-         \hp3&\ \ \quad -\hp2&\ \ \quad -         \hp2&\ \ \quad -     \hp1&\ty 51\er10$^\dagger$    \hp3\\[-0.5ex]
 &$\PN(2000)5/2^+$    & 11\er5     \hf3&   9\er4  & 1  &     16\er4 & 3  &\ \ \quad -        \hf1& \ \ 2\er 2 \hf2&\ \ \quad -          \hf2& \ \ 28\er9\hf1& 28\er 15\hf2\\[-1ex]
                   && \ty8\er  4\hp0&\ty22\er10 &    &\ty  34\er15&    &\ \ \quad -        \hp1& \ \ \ty21\er10\hp2&\ \ \quad -        \hp2& \ \ \ty16\er9\hp1&\ty 10\er 5\hp0\\[-0.5ex]
 &$\PN(1990)7/2^+$    &1\er1  \hf3&   65\er20 & 3  &     -      & 5  & \ \ $<$2      \hf1& \ \ $<$2    \hf1& \ \ $<$2        \hf4&\ \ \quad -    \hf1& \quad -\hf4\\[-1ex]
                   &&\ty1.5\er0.5\hp0&\ty48\er10&    &     -      &    & \ \ \ty$<$2   \hp1& \ \ \ty$<$2 \hp1& \ \ \ty$<$2     \hp4&\ \ \quad -    \hp1& \quad -\hp1\\[-0.5ex]
 \hline\hline
\parbox[t]{3mm}{\multirow{10}{*}{\rotatebox[origin=c]{90}{3rd shell, mixed oscillations}}}
& $\PN(1895)1/2^-$    & 6\er4\hf0    &    x       &    &      5\er 3& 2  & \ \ 2\er 2    \hf0&\ \ \quad -        \hf1&\ \ \quad -       \hf1&\ \ \quad -     \hf2& 11\er7 \hf1\\[-1ex]
                   && \ty2.5\er1.5\hp0&         &    &\ty   7\er 4&    & \ \ \ty8\er 8 \hp0&\ \ \quad -        \hp1&\ \ \quad -       \hp1&\ \ \quad -     \hp2&\ty18\er15\hp0\\[-0.5ex]
& $\PN(1875)3/2^-$    & 3\er 2   \hf2&6\er4      & 0  &      4\er 3& 2  & \ \ 11\er5     \hf2& \ \ 4\er4      \hf1& \ \ 2\er2     \hf1&\ \ \quad -     \hf2&50\er20 \hf1\\[-1ex]
                   && \ty4\er 2\hp0&\ty14\er7   &    &\ty   7\er 5&    & \ \ \ty5\er3  \hp2& \ \ \ty$<2$   \hp1& \ \ \ty$<$1  \hp1&\ \ \quad -     \hp2&\ty45\er15\hp0\\[-0.5ex]
&$\PgD(1900)1/2^-$& 4\er 4   \hf0&    x       &    &     60\er25 & 2  & \ \ 15\er12    \hf0& \ \ 10\er7     \hf1&\ \ \quad -       \hf1&\ \ \quad -     \hf2&\quad x      \hp0\\[-1ex]
                   && \ty7\er 2\hp0&            &    &\ty  50\er20&    & \ \ \ty20\er12\hp0& \ \ \ty6\er4  \hp1&\ \ \quad -       \hp1&\ \ \quad -     \hp2&\\[-0.5ex]
&$\PgD(1940)3/2^-$& 13\er6  \hf0&16\er6     & 0  &     30\er12 & 2  & \ \ 5\er5     \hf2& \ \ 7\er5     \hf1& \ \ 20\er13    \hf1&\ \ \quad -     \hf2&\quad x      \hp0\\[-1ex]
                   && \ty2\er 1\hp0&\ty46\er20  &    &\ty  12\er 7&    & \ \ \ty7\er7  \hp2& \ \ \ty4\er3  \hp1& \ \ \ty8\er6 \hp1&\ \ \quad -     \hp2&\\[-0.5ex]
& $\PN(2120)3/2^-$    & 5\er 2   \hf2&10\er5     & 0  &     14\er7  & 2  & \ \ 6\er6     \hf2& \ \ 10\er5   \hf1& \ \ $<$2   \hf1&\ \ \quad -     \hf2&4\er3\hf1\\[-1ex]
                   && \ty5\er 3\hp0&\ty50\er20  &    &\ty  20\er12&    & \ \ \ty10\er10\hp2& \ \ \ty15\er10\hp1& \ \ \ty15\er8\hp1&\ \ \quad -     \hp2& \ty11\er4\hp0\\[-0.5ex]
& $\PN(2060)5/2^-$    &11\er 2   \hf2&9\er3        & 2  &     -      & 4  & \ \ 8\er 5\hf2& \ \ 13\er 7   \hf1&\ \ \quad -       \hf3& \ \ 14\er5 \hf0&5\er3\hf3\\[-1ex]
                   &&\ty11\er 2\hp0&\ty7\er3    &    &     -      &    & \ \ \ty 9\er 5\hp2& \ \ \ty15\er 6\hp1&\ \ \quad -       \hp3& \ \ \ty15\er7\hp0&\ty6\er3\hp0\\[-0.5ex]
& $\PN(2190)7/2^-$    &15\er 3   \hf4&4\er 2     & 2  &     -      & 4  & \ \ \quad-    \hf4& \ \ \quad-    \hf3&\ \ \quad -       \hf3&\ \ \quad -     \hf2& 6\er3\hf3\\[-1ex]
                   &&\ty16\er 2\hp0&\ty25\er 6  &    &     -      &    & \ \ \quad-    \hp4& \ \ \quad-    \hp3&\ \ \quad -       \hp3&\ \ \quad -     \hp2&\ty5\er3\hp0\\[-0.5ex]
\hline

\end{tabular}}
\renewcommand{\arraystretch}{1.0}\vspace{-3mm}
\end{center}
 \end{table*}
As mentioned in the introduction and
  discussed in further detail in \cite{Sokhoyan:2015fra}, using a simple quark model
  picture and the respective wave functions,
 baryon excitations can be divided into classes in which the excitation energy
fluctuates between the $\rho$ and $\lambda$ oscillator (one-oscillator excitations) and in which,
in part of their wave function, both oscillators are excited simultaneously (mixed-oscillator
excitations). In \cite{Thiel:2015xba} it was argued that mixed-oscillator excitations prefer
to de-excite in a two-step process in which first one oscillator releases its energy and than the
second oscillator. Table~\ref{tab:Decayall} is separated into different blocks: the decay modes
into $\PN\pi$ and $\PgD\pi$ have final states without excitation (the $\PgD(1232)$ is the
ground state of the baryon decuplet). In the other decay modes, the intermediate resonances carry spatial
excitation, either the baryon or the meson (the $\sigma$ or $f_0(500)$). In addition, the states in the table are sorted by
the kind of baryon excitation: one-oscillator or mixed-oscillator excitations.
All one-oscillator excitations have a significantly higher branching ratio into the ground states $\PN\Pgp$ or $\PgD\Pgp$, than into the excited states $\PN(1520)\pi$, $\PN(1535)\pi$, or $\PN\sigma$, as depicted in Fig.~\ref{fig:BRs}: the black dots compared to the red squares. They decay on average with about $\unit[(60\pm3)]{\%}$ into $\PN\Pgp$ or $\PgD\Pgp$ and with about $\unit[(6\pm1)]{\%}$ into the aforementioned excited states. The residual $\unit[(34\pm3)]{\%}$ are assigned to other two and multiparticle final states like, \textit{e.g.} $\PgD\Pgo$.
In contrast, resonances with mixed-oscillator excitations have branching ratios for decays into both $\PN\Pgp/\PgD\Pgp$ and into the excited states as shown in Tab.~\ref{tab:Decayall} (blue dots and green squares in Fig.~\ref{fig:BRs}): on average about $\unit[(30\pm3)]{\%}$ go to $\PN\Pgp$ or $\PgD\Pgp$ and about $\unit[(17\pm2)]{\%}$ go to orbitally excited states.

\begin{figure}
\resizebox{.5\textwidth}{!}{%
  \includegraphics{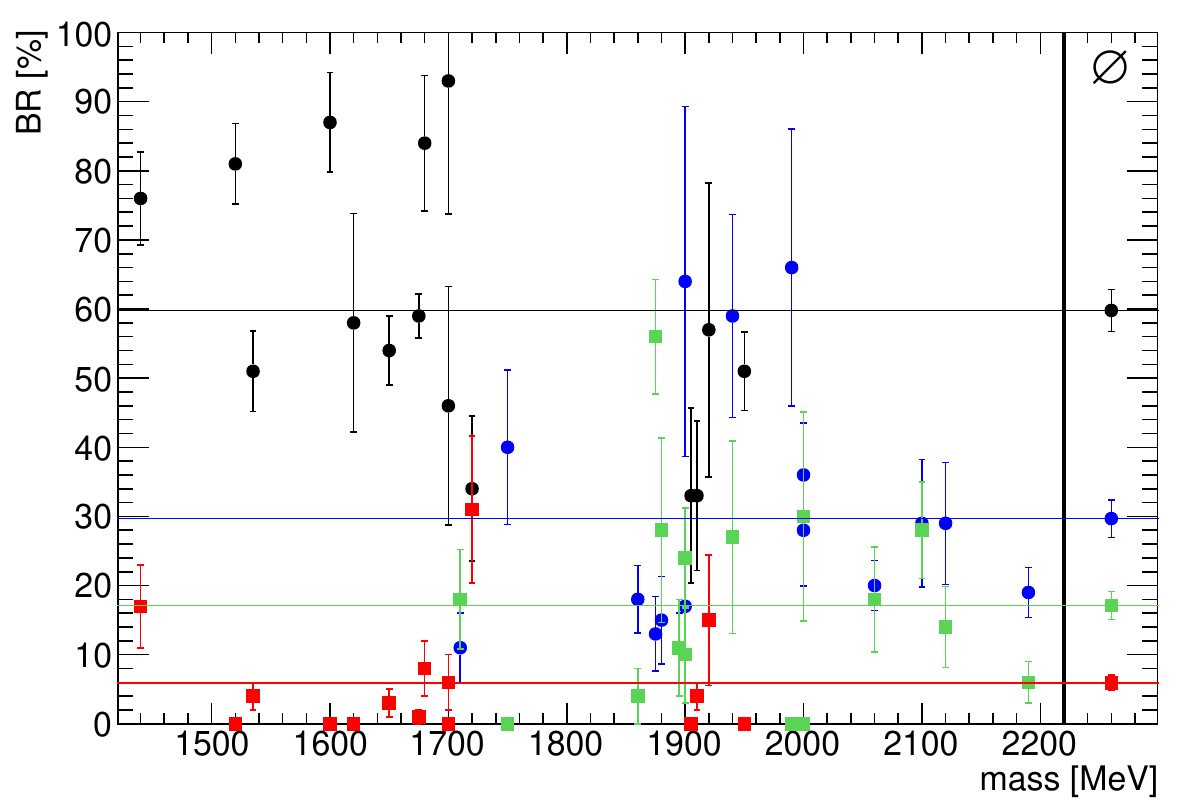}
} \caption{The sums of branching ratios into $\PN\Pgp$, $\PgD\Pgp$ are compared to the sums of branching ratios into $\PN(1520)\Pgp$, $\PN(1535)\Pgp$, and $\PN\sigma$ of the resonances listed in Tab.~\ref{tab:Decayall}. The branching ratios (BR) of one-oscillator excitations decaying into the ground states $\PN\Pgp$ or $\PgD\Pgp$ are depicted as black dots, when decaying into the excited states $\PN(1520)\Pgp$, $\PN(1535)\Pgp$, or $\PN\sigma$ as red squares. Mixed-oscillator states decaying into the ground states mentioned above are shown as blue dots, when decaying into the excited states mentioned above as green squares. Additionally the mean branching ratios are shown on the right as well as colored lines.}
\label{fig:BRs}
\end{figure}

This finding still holds when restricted to resonances of similar mass and same quantum numbers, namely the four $\PN^*$ and four $\PgD^*$ resonances of positive parity with a mass around \unit[1900]{MeV}. Here, the $\PgD^*$ resonances have one excited oscillator and decay into $\PN\Pgp/\PgD\Pgp$ with an average branching ratio of about $\unit[(44\pm7)]{\%}$, while their average branching ratio to the excited states mentioned above, is about $\unit[(5\pm2)]{\%}$. On the other hand, the $\PN^*$ resonances, which have mixed-oscillator excitations, decay into $\PN\Pgp/\PgD\Pgp$ with an average branching ratio of about $\unit[(34\pm6)]{\%}$ and into the orbitally excited states with about $\unit[(21\pm5)]{\%}$, similar to the findings without the mass restriction.

The new branching ratios are mostly compatible with the ones determined in Ref.~\cite{Sokhoyan:2015fra}.
In some cases, the new branching ratios are significantly smaller compared to the old ones. This is likely due to the interference between two resonances with the same spin-parity but different isospin.
These two contributions were not properly identified in $\Pgg\Pp\to\Pp\Pgpz\Pgpz$ but are under control when data with two charged pions are included in the data base. 

\section{Conclusion/Summary}
\label{sec:summary}
We have reported new data on polarization and double-polarization observables for the reaction
$\Pgg \Pp\to \Pp\Pgpz\Pgpz$. The data were included in the BnGa PWA which contains a large data base 
on pion- and photo-induced reactions off protons and neutrons \cite{bnga_website}. The consistency with earlier results is good. In particular, we confirm the
conjecture that resonances in which, based on a quark model picture, a fraction of the wave function carries a simultaneous
excitation of both oscillators have a significant probability to decay by sequential decays, first into
an excited meson or baryon resonance, then into the three-particle final state.

\begin{acknowledgement}
This paper is dedicated to the memory of V.~Nikonov.\bigskip

We thank the technical staff at ELSA and at all the participating institutions for their
invaluable contributions to the success of the experiment. Funded by the 
Deutsche Forschungsgemeinschaft (DFG, German Research Foundation)  Project-ID 196253076  TRR 110,
the U.S. Department of Energy, Office of Science, Office of Nuclear Physics, under Contract No. DE-FG02-92ER40735,
and the Schweizerische Nationalfonds (200020-175807, 156983, 132799, 121781, 117601). The
work was also partially supported by the EU Horizon 2020 research and innovation
program, STRONG-2020 project, under grant agreement No. 824093.

Part of this work comprises part of the PhD thesis of T.~Seifen \cite{diss:seifen}.
\end{acknowledgement}



\begin{thebibliography}{}
\bibitem{Capstick:2000qj}
  S.~Capstick and W.~Roberts,
  Prog.\ Part.\ Nucl.\ Phys.\  {\bf 45} (2000) S241.

    \bibitem{Capstick:1986bm}
  S.~Capstick and N.~Isgur,
  Phys.\ Rev.\  D {\bf 34} (1986) 2809.

 \bibitem{Loring:2001kx}
  U.~L\"oring, B.~C.~Metsch and H.~R.~Petry,
  Eur.\ Phys.\ J.\  A {\bf 10} (2001) 395.

\bibitem{anselmino}
  M.~Anselmino, E.~Predazzi, S.~Ekelin, S.~Fredriksson and D.~B.~Lichtenberg, Rev.\ Mod.\ Phys.\  {\bf 65} (1993) 1199.
    
\bibitem{RPP}
S. Navas et al. (Particle Data Group), Phys. Rev. D 110, 030001 (2024)


\bibitem{Anisovich:2017bsk}
A.~V.~Anisovich, V.~Burkert, M.~Had\v{z}imehmedovi\'c, D.~G.~Ireland, E.~Klempt, V.~A.~Nikonov, R.~Omerovi\'c, H.~Osmanovi\'c, A.~V.~Sarantsev and J.~Stahov, \textit{et al.}
Phys. Rev. Lett. \textbf{119} (2017) no.6, 062004

\bibitem{Edwards:2011jj}
  R.~G.~Edwards {\it et al.}, 
  Phys.\ Rev.\ D {\bf 84} (2011) 074508.
    
\bibitem{Khan:2020ahz}
T.~Khan, D.~Richards and F.~Winter,
Phys. Rev. D \textbf{104} (2021) no.3, 034503. 

\bibitem{Eichmann:2016yit}
G.~Eichmann, H.~Sanchis-Alepuz, R.~Williams, R.~Alkofer and C.~S.~Fischer,
Prog. Part. Nucl. Phys. \textbf{91} (2016), 1-100

\bibitem{Eichmann:2018adq}
G.~Eichmann and C.~S.~Fischer,
Few Body Syst. \textbf{60} (2019) no.1, 2

\bibitem{Qin:2019hgk}
S.~x.~Qin, C.~D.~Roberts and S.~M.~Schmidt,
Few Body Syst. \textbf{60} (2019) no.2, 26

\bibitem{Barabanov:2020jvn}
M.~Y.~Barabanov \textit{et al.},
Prog. Part. Nucl. Phys. \textbf{116} (2021), 103835

\bibitem{Lutz:2005ip}
  M.~F.~M.~Lutz and E.~E.~Kolomeitsev,
  Nucl.\ Phys.\ A {\bf 755} (2005) 29.
  
  \bibitem{Guo:2017jvc}
  F.~K.~Guo, C.~Hanhart, U.~G.~Mei\ss ner, Q.~Wang, Q.~Zhao and B.~S.~Zou,
  Rev.\ Mod.\ Phys.\  {\bf 90}, no. 1 (2018) 015004.

\bibitem{Mai:2022eur}
M.~Mai, U.~G.~Mei\ss{}ner and C.~Urbach,
Phys. Rept. \textbf{1001} (2023), 1-66.

\bibitem{Crede:2013kia}
V.~Crede and W.~Roberts,
Rept. Prog. Phys. \textbf{76} (2013), 076301

\bibitem{Thiel:2022xtb}
A.~Thiel, F.~Afzal and Y.~Wunderlich,
Prog. Part. Nucl. Phys. \textbf{125} (2022), 103949

\bibitem{EK2025}
V. Burkert, G. Eichmann, E. Klempt, 
``The impact of $\gamma N$ and $\gamma^* N$ interactions on our understanding of nucleon excitations,'' in preparation 



\bibitem{Glozman:1999tk}
  L.Y.~Glozman,
  Phys.\ Lett.\  B {\bf 475} (2000) 329.

\bibitem{Jaffe:2004ph}
R.L.~Jaffe,
  Phys.\ Rept.\  {\bf 409} (2005) 1.
 \bibitem{Glozman:2007ek}
  L.Y.~Glozman,
  Phys.\ Rept.\  {\bf 444} (2007) 1.

\bibitem{Afonin:2007mj}
    S.S.~Afonin,
  Int.\ J.\ Mod.\ Phys.\  A {\bf 22} (2007) 4537.

\bibitem{Shifman:2007xn}
 M.~Shifman, A.~Vainshtein,
  Phys.\ Rev.\  D {\bf 77} (2008) 034002.

    \bibitem{Klempt:2002tt}
  E.~Klempt,
  Phys.\ Lett.\  B {\bf 559} (2003) 144.

    \bibitem{Forkel:2008un}
  H.~Forkel and E.~Klempt,
  Phys.\ Lett.\  B {\bf 679} (2009) 77.

\bibitem{Brodsky:2014yha}
S.~J.~Brodsky, G.~F.~de Teramond, H.~G.~Dosch and J.~Erlich,
Phys. Rept. \textbf{584} (2015), 1-105.

\bibitem{Klempt:2002vp}
  E.~Klempt,
  Phys.\ Rev.\  C {\bf 66} (2002) 058201.

\bibitem{Anisovich:2015gia} 
  A.~V.~Anisovich {\it et al.},
  Phys.\ Lett.\ B {\bf 766} (2017) 357.

\bibitem{Sokhoyan:2015fra}
  V.~Sokhoyan {\it et al.},
  Eur.\ Phys.\ J.\ A {\bf 51} (2015) 95,
  Erratum: [Eur.\ Phys.\ J.\ A {\bf 51} (2015) 187].

\bibitem{Thiel:2015xba}
  A.~Thiel {\it et al.}  [CBELSA/TAPS Collaboration],
  Phys.\ Rev.\ Lett.\  {\bf 114} (2015) 091803.

\bibitem{elsa}
 W.~Hillert, Eur.\ Phys.\ J.\ A {\bf 28S1} (2006) 139.
\bibitem{elsner}
D.~Elsner {\it et al.} 
, Eur.\ Phys.\ J.\ A {\bf 33} (2007) 147.
\bibitem{bradtke}
C.~Bradtke {\it et al.}, Nucl.\ Instrum.\ Meth.\ A {\bf 436} (1999) 430.
\bibitem{cb}
 E.~Aker {\it et al.} 
 , Nucl.\ Instrum.\ Meth.\ A {\bf 321} (1992) 69.
\bibitem{novotny}
R.~Novotny 
, IEEE Trans.\ Nucl.\ Sci.\  {\bf 38} (1991) 379.
\bibitem{gabler}
A.~R.~Gabler {\it et al.}, Nucl.\ Instrum.\ Meth.\ A {\bf 346} (1994) 168.
\bibitem{suft}
G.~Suft {\it et al.}, Nucl.\ Instrum.\ Meth.\ A {\bf 538} (2005) 416.
\bibitem{gottschall}
M.~Gottschall {\it et al.}, Eur.\ Phys.\ J.\ A {bf 57} (2021) 40
\bibitem{novosibirsk}
H.~Ikeda {\it et al.} 
, Nucl.\ Instrum.\ Meth.\ A {\bf 441} (2000) 401.
\bibitem{pee}
H.~van~Pee {\it et al.}, Eur.\ Phys.\ J.\ A {\bf 31} (2007) 61.
\bibitem{williams}
M.~Williams, M.~Bellis and C.~A.~Meyer, JINST {\bf 4} (2009) P10003.
\bibitem{roberts}
W.~Roberts and T.~Oed, Phys.\ Rev.\ C {\bf 71} (2005) 055201.
\bibitem{barker}
I.~S.~Barker, A.~Donnachie and J.~K.~Storrow, Nucl.\ Phys.\ B {\bf 95} (1975) 347.
\bibitem{maid}
A.~Fix and H.~Arenh\"ovel, Eur.\ Phys.\ J.\ A {\bf 25} (2005) 115
\bibitem{bnga}
E.~Gutz {\it et al.} 
, Eur.\ Phys.\ J.\ A {\bf 50} (2014) 74.

\bibitem{Doring:2025sgb}
M.~D\"oring, J.~Haidenbauer, M.~Mai and T.~Sato,
``Dynamical coupled-channel models for hadron dynamics,''
[arXiv:2505.02745 [nucl-th]].

\bibitem{Anisovich:2004zz}
  A.~Anisovich, E.~Klempt, A.~Sarantsev and U.~Thoma,
  Eur.\ Phys.\ J.\ A {\bf 24} (2005) 111.

\bibitem{Anisovich:2006bc}
  A.~V.~Anisovich and A.~V.~Sarantsev,
  Eur.\ Phys.\ J.\ A {\bf 30} (2006) 427.

\bibitem{Anisovich:2007zz}
  A.~V.~Anisovich, V.~V.~Anisovich, E.~Klempt, V.~A.~Nikonov and A.~V.~Sarantsev,
  Eur.\ Phys.\ J.\ A {\bf 34} (2007) 129.

\bibitem{Denisenko:2016ugz}
  I.~Denisenko {\it et al.},
  Phys.\ Lett.\ B {\bf 755} (2016) 97.

\bibitem{Arndt:2006bf}
  R.A.~Arndt {\it et al.}, 
  Phys.\ Rev.\  C {\bf 74} (2006) 045205.

\bibitem{Hohler:1979yr}
  G.~H\"ohler, F.~Kaiser, R.~Koch and E.~Pietarinen,
  ``Handbook Of Pion Nucleon Scattering,''
 Fachinform. Zentr. Karlsruhe 1979, 440 P. (Physics Data, No.12-1 (1979)).

\bibitem{mueller}
  J.~M{\"u}ller {\it et al.} 
  , Phys.\ Lett.\ B {\bf 803} (2020) 135323.

\bibitem{Golovatch:2018hjk}
  E.~Golovatch {\it et al.} 
  , Phys.\ Lett.\ B {\bf 788} (2019) 371.

\bibitem{Hunt}
  B.~C.~Hunt and D.~M.~Manley
  , Phys.\ Rev.\ C {\bf 99} (2019) 055205.

\bibitem{bnga_website}
  \url{pwa.hiskp.uni-bonn.de}
  
\bibitem{diss:seifen}
 T.~Seifen, PhD thesis,
  \url{https://nbn-resolving.org/urn:nbn:de:hbz:5-61416}

\end{thebibliography}
\end{document}